\documentclass[twocolumn,amssymb, amsmath, showpacs,pra]{revtex4}
\usepackage{epsfig,amsmath,graphicx,amssymb}
\usepackage[colorlinks=fals,linkcolor=blue]{hyperref}
\usepackage{graphicx}
\usepackage{dcolumn}

\usepackage{epsfig}
\usepackage{color}

\usepackage{fancyhdr}
\usepackage{latexsym}
\usepackage{epsfig}
\usepackage{pstcol}
\usepackage{pst-text}
\usepackage{pst-node}
\usepackage{pstricks}
\usepackage{shadow,epsf,amsthm,amssymb,amsmath}
\usepackage{enumerate}

\begin{document}
\title{Geometric Manipulation of Ensembles of Atoms on Atom Chip for Quantum Computation}
\author{Yicong Zheng}

\email{
yicongzh@usc.edu
}
\author{Todd A. Brun}
\email{tbrun@usc.edu}
\affiliation{Department of Electrical Engineering, Center for Quantum Information Science \& Technology, University of Southern California, Los Angeles, California, 90089\\}
\date{\today}

\begin{abstract}
We propose a feasible scheme
to achieve quantum computation
based on geometric manipulation of ensembles of atoms, and
analyze it for neutral rubidium atoms magnetically trapped in
planoconcave microcavities on an atom chip. The geometric
operations are accomplished by optical excitation of a single
atom into a Rydberg state in a constant electric field. Strong
dipole-dipole interactions and incident lasers drive the dark
state of the atom ensembles to undergo cyclic
evolutions that realize a universal set of quantum gates. Such
geometric manipulation turns out naturally to protect the qubits
from the errors induced by non-uniform laser illumination as well
as cavity loss. The gate performance and decoherence processes
are assessed by numerical simulation.

\end{abstract}
\maketitle


\section{Introduction}
Quantum information processing (QIP) holds out the possibility to run algorithms
 and protocols superior to those of its
classical counterpart~\cite{Bennett,Nielson&Chuang}. In recent
decades, numerious candidates for physical implementation of quantum
information processors have been proposed \cite{Quantum Computing
Devices}. Because of long coherence times and exceptional controlability,
quantum optical and atomic systems such as trapped ions \cite{Ions
trap_1, Ions_trap_2}, neutral atoms \cite{Neutral_atom,
Neutral_atom2}, and cavity QED \cite{Cavity_QED, Cavity_QED_Guo},
have taken a leading role in implementing quantum logic. The photon is a
remarkably robust candidate for a qubit, and easy to transport  long
distances; however, as we know, they interact only weakly with
each other, which makes the realization of quantum gates based on
photons difficult. For instance, in linear optical quantum
computation(LOQC) \cite{LOQC}, extra measurements are required.
On the other hand, ensembles of trapped atoms or molecules may serve
as convenient and robust quantum memories for photons. They can act as an
interface with flying qubits, and
store and retrieve single photons, for example by electromagnetically induced
transparency (EIT) \cite{Lukin_EIT_Review, EIT_Review}, which allows to keep
coherence of quantum states up to several hundreds of millisecond \cite{EIT_Experiment_2009}.

One might therefore hope
to use controlled interactions in the ensemble of atoms, once the
flying qubits are stored, to realize
universal quantum logic gates in a deterministic
and scalable way. Several promising schemes for such operations based on combining EIT with Rydberg
 atoms was proposed in Ref. \cite{MDLukin2001, Photon_Photn_Interaction_2011}, exciting atoms with lasers to high-lying Rydberg states
 and exploiting the strong long-range dipole-dipole interaction between Rydberg
 states. Recent theoretical proposals have extended towards a single step high-fidelity entanglement of a mesoscopic number of atoms \cite{Muller2009} and quantum simulation both coherent dynamics and dissipative evolution process for many body system \cite{Weimer2010}. Remarkably, the building blocks behind all such setup-up have
 been demonstrated experimentally by several groups \cite{Gaetan2009, Urban2009}.
In Ref. \cite{Petresyan}, an effective strong
dipole-dipole blockade effect is achieved by coupling to
microwave coplanar waveguide resonators to realize quantum logic on
ensembles of molecules. Here, we put forward an
alternative, hopefully scalable approach to realize universal
quantum gates based solely on laser-controlled geometric manipulations
of neutral atomic clouds trapped on the surface of atom chips.
The use of geometric gates can naturally eliminate certain kinds of errors.

The paper is organized as follows. In section II we introduce the basic model of atoms and cavity system.  Then, we show how to construct the Hamiltonian of the system and the method of adiabatically manipulating the system evolution to implement three quantum gates geometrically: the phase gate, the y-rotation gate, and the controlled phase gate.  These constructions are given in sections IIIA, IIIB, and IIIC , respectively. Numerical analysis of their performance is also given in these three sections.  These three gates forms a well-known universal set of quantum logical gates. Finally, in section IV, we discuss how to perform qubit measurements on this system, and conclude in section V.

\section{Model}
\begin{figure}[htp1]
\centering\includegraphics[width=80mm,height=60mm]{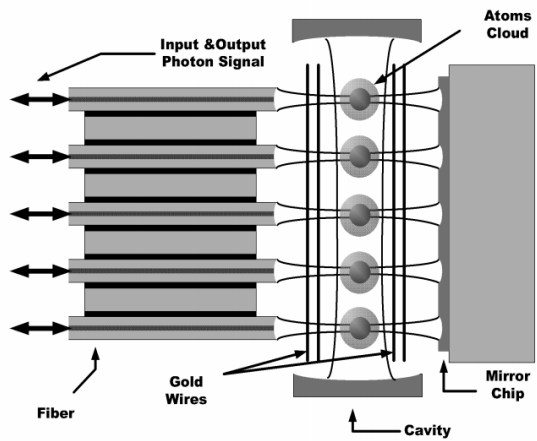}
\caption{\label{Fig:scheme} Schematic of trapped ensembles of
atoms in magnetic atom traps inside plano-concave optical microcavities (horizontal).
Each atomic cloud is a qubit, and forms
a processing cell. Different qubits couple to each other mediated by the
photonic modes of another Fabry-Perot cavity (vertical). }
\end{figure}

Atom chips supply a good platform on which precise control and
manipulation of the neutral atoms can be performed
\cite{Folman2002,Review_of_Modern_Physics_Atomschip}.
We will show that an atom chip with integrated Fabry-Perot (FP) microcavity can realize
a universal set of all-optical control universal quantum gates for atomic ensemble qubits.

Consider several plano-concave Fabry-Perot (FP) microcavity
resonators \cite{Trupke2007} integrated on the
atom chip, which are parallel to each other.
Another large FP cavity is integrated so that its mode is perpendicular to all the
plano-concave cavities. The diagram is shown in Fig.~\ref{Fig:scheme}.
The plano-concave resonator
consists of an isotropically etched dip in a silicon surface, and
the cleaved tip of a single-mode fibre \cite{Trupke2005} serves as the input-output channel.
Atomic ensembles are placed in the region of highest field strength of
the cavity modes, leading to higher values of the coupling $g$, and a $Q$
value over $10^{6}$. The perpendicular FP cavity modes serve as a
data bus to couple between  different qubits.

\begin{figure}[htp1]
\centering\includegraphics[width=70mm,height=50mm]{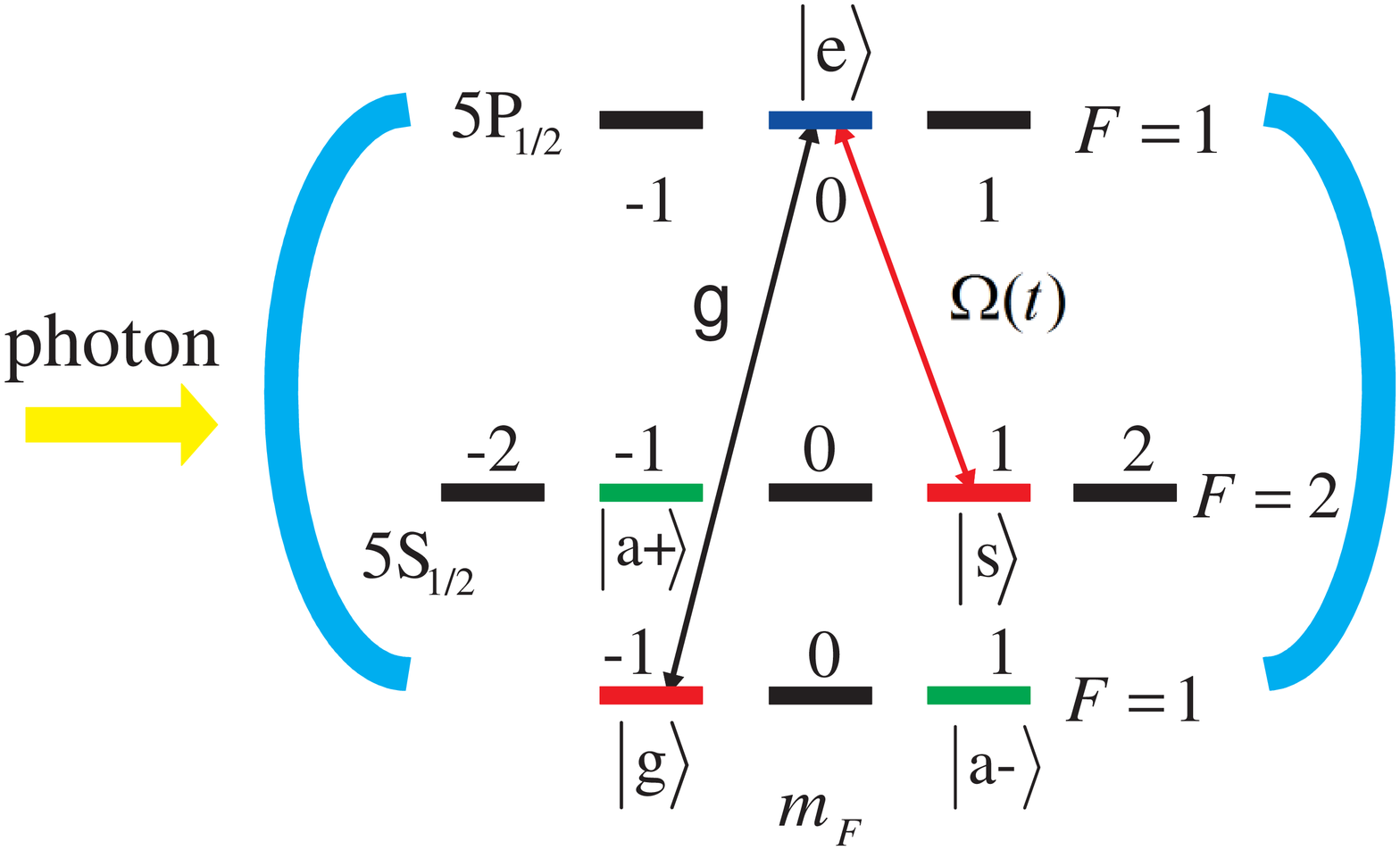}
\caption{\label{Fig:atom level} Storage of single photon qubit. Level
scheme of a single $^{87}\textrm{Rb}$ atom is shown. The input qubit is encoded
as $|0\rangle$ if there is no photon in the input mode and
$|1\rangle$ if there is a photon in the input mode. Here, we use the hyperfine levels
of manifold 5$\textrm{S}_{1/2}$: clock states $|g\rangle=|F=1,m_F=-1\rangle$ and
$|s\rangle=|F=2,m_F=1\rangle$; $|a_{-}\rangle=|F=1,m_F=1\rangle$
and $|a_{+}\rangle=|F=2,m_F=-1\rangle$ are used as ancillary states.
And a state of manifold 5$\textrm{P}_{1/2}$ serves as an intermediate state
$|e\rangle$ $(|F=1,m_F=0\rangle)$.
The storage process mapping single photon signals into
collective excited atomic states of $|s\rangle$
can be accomplished by adiabatically turning on the classical laser
field $\Omega(t)$ to a value much larger than the coupling
constant $g$ when the photon arrives. The reverse process can be
used to read out the photon.}
\end{figure}

The atomic clouds must be confined in traps
inside the cavities. We can introduce  magnetic fields produced by
current-carrying wires on the surface and coupled to the magnetic dipole
of the atoms. The small scale of the structure produces strong magnetic
field gradients, which make tight traps for magnetic atoms. Atoms in a
weak-field-seeking state will be attracted and held in this region.
The hyperfine states $|F=2,m_F=1\rangle$ and $|F=1,m_F=-1\rangle$ of
$5\textrm{S}_{1/2}$ of  $^{87}\textrm{Rb}$ are ideal weak-field-seeking
states that can be trapped by a static magnetic potential \cite{Folman2002}.
In addition, these two states have opposite Land\'{e} factors, so that
they experience identical magnetic potentials. Thus, the decoherence of an
arbitrary superposition of these two states due to current intensity fluctuations is
strongly inhibited \cite{Phase_gate_atom chips}. Coherent oscillations
between these states have been observed with decoherence
time as long as $\tau_c=2.8\pm1.6$s \cite{Treutlein2004}.
So, it is natural to consider these two states of $^{87}\textrm{Rb}$ for
quantum information processing.

The state of the $^{87}\textrm{Rb}$ atom cloud can be initialized by inputting a
single photon qubit to the plano-concave cavity through the fibre,
as shown in Fig.~\ref{Fig:atom level}. The state $|0\rangle_\textrm{ph}$ corresponds to
no photons in the input channel, and the state
$|1\rangle_{\textrm{ph}}$ corresponds to one photon. An arbitrary
qubit state can be represented as $\alpha|0\rangle_{\textrm{ph}}
+ \beta|1\rangle_{\textrm{ph}}$. In general, the  state of $n$ input
channels is an entangled state in a Hilbert space of dimension $2^n$.

An ensemble of $N$ identical multi-state atoms is trapped in each
cell. Using well-developed techniques, all atoms can be initially prepared and
trapped in a specific sub-level (hyperfine level $|g_i\rangle$, $i = 1,2,3,...,N$)
of the internal atomic ground space of manifold \cite{MDLukin2001}.
Relevant states of each atom include the other clock state $|s_i\rangle$,
two ancillary states $|a_{-i}\rangle, |a_{+i}\rangle$, and an intermediate
state $|e_i\rangle$. (As mentioned earlier, $|s_i\rangle$ has a long coherence time.)
We also assume the atomic density is not too high, so that interaction
between atoms can be safely neglected.
Atoms are manipulated by illuminating the entire ensemble,
to excite all atoms with equal probability, so only symmetric
collective states are involved in this process \cite{MDLukin2001}. In
the case of loading quantum information into the ensembles, we
consider only three states: $|e_i\rangle$, $|s_i\rangle$ and
$|g_i\rangle$, where $|e_i\rangle$ and $|g_i\rangle$ are coupled to
the cavity mode of the plano-concave cavity, while $|e_i\rangle$ and
$|s_i\rangle $ are coupled by a classical field. Initially, all the
atoms are in their ground states:
\[
|\textbf{g}\rangle = |\textbf{g}^N\rangle = |g_1\rangle...|g_N\rangle .
\]
We define operators
\[
\mathcal {\hat{S}} = \frac{1}{\sqrt{N}}\sum_{i}|g_i\rangle\langle s_i| ,
\]
\[
\mathcal{\hat{E}} = \frac{1}{\sqrt{N}}\sum_{i}|g_i\rangle\langle e_i| ,
\]
\[
\mathcal {\hat{A}_\pm} = \frac{1}{\sqrt{N}}\sum_{i}|g_i\rangle\langle a_{\pm i}| .
\]
The storage state with excitation $n$ is
\[
|\textbf{s}^n\rangle \equiv |\textbf{g}^{N-n} , \textbf{s}^n\rangle
  = \sqrt{\frac{(N-n)!N^n}{N!n!}}(\mathcal{\hat{S^\dag}})^n|\textbf{g}\rangle .
\]
For the case when $n=1$, this is
\begin{eqnarray*}
|\textbf{s}\rangle &=& \frac{1}{\sqrt{N}}\bigl(|s_1,g_2,g_3...g_N\rangle +
|g_1,s_2,g_3...g_N\rangle \\
&& + \cdots + |g_1,g_2,g_3...s_N\rangle\bigr) .
\end{eqnarray*}
The commutation relation of $\mathcal {\hat{S}}$ and
$\mathcal{\hat{S}^\dag}$ is
\[
[\mathcal {\hat{S}}, \mathcal {\hat{S}^\dag}] = 1-\frac{2n}{N} .
\]
For a sufficiently small number of excitations $(n
\ll N)$, these two operators can be approximated by bosonic annihilation
and creation operators of quasiparticles corresponding to $|s\rangle$. Operators $\mathcal
{\hat{E}}$ and $\mathcal {\hat{E}^\dag}$, and $\mathcal {\hat{A}_\pm}$ and $\mathcal
{\hat{A}_\pm^\dag}$ have the same properties.

 As shown in Ref. \cite{Trap_photon} and
Ref. \cite{Trap_correlated_photon}, when the photon arrives in the
cavity, one can adiabatically turn on the classical field coupling the states $|e\rangle$ and $|s\rangle$, until the Rabi
frequency $\Omega$ is much larger than the cavity-atom coupling constant
$g$, so that the state of photon is stored in the
ensemble of atoms in the cavity in the form $|n\rangle_{\textrm{ph}}
\rightarrow |\textbf{s}^n\rangle$, where $|n\rangle_{\textrm{ph}}$ represents
the $n$-photon Fock state. This process is reversible: one can
retrieve the photon from the ensemble of atoms. So, an arbitrary input
state $\alpha|0\rangle_{\textrm{ph}} + \beta|1\rangle_{\textrm{ph}}$ can be mapped
into the state $\alpha|\textbf{g}\rangle + \beta|\textbf{s}\rangle$ with
high fidelity \cite{Trap_photon}. Then, We denote $|0\rangle_L =
|\textbf{g}\rangle$ and $|1\rangle_L =
|\textbf{s}\rangle$ for the logical representation of a qubit in each cloud of atoms.

\section{A Universal Set of Geometric Gates}

We now discuss how to geometrically implement quantum computation
(holonomic quantum computation \cite{Zanardi}) in our system. The
first step is to construct a Hamiltonian that has an eigenspace with
eigenvalue 0 to avoid dynamic phase during the cyclic evolution.
This eigenspace can either be nondegenerate, which
would introduce simple Abelian phase factors (Berry
phases) \cite{Berry} or denerate, which could cause general non-Abelian
operations \cite{Wilczek}. It is well known that single-qubit
operations, together with a nontrivial two-bit gate, make a
universal set of quantum logic gates for quantum computation \cite{Llyod}.
By constructing an appropriate set of looped paths in the parameter
space we can obtain an arbitrary unitary transformation in the computational space.

We choose for our universal set of gates
\[
R_z^{(i)}(\phi_1) = \exp\biggl(i\phi_1|1_i\rangle_{LL}\langle1_i|\biggr) ,
\]
\[
R_y^{(i)}(\phi_2) = \exp\biggl(i\phi_2\sigma_i^y\biggr) ,
\]
\[
U^{(jk)}(\phi_3) = \exp\biggl(i\phi_3|1_j0_k\rangle_{LL}\langle1_j0_k|\biggr) .
\]
The states $|0_i\rangle_{L}$ and $|1_i\rangle_{L}$ are the computational
basis states for qubit $i$, and $\sigma_i^y$ is the Pauli operator in the $y$
direction for the $i$th qubit. It is well know that if we can implement these gates
with arbitrary $\phi_1, \phi_2, \phi_3$, we can implement all unitary
transformations \cite{Llyod}. The three operators correspond to a rotation
in the $z$ direction, a rotation in  the $y$ direction, and a conditional phase
rotation when qubits $j$ and $k$ are in the state $|10\rangle_L$. In this section,
we show how to holonomically realize these three gates.

\subsection{One Qubit Phase Gate}

\begin{figure}[htp1]
\centering\includegraphics[width=75mm,height=40mm]{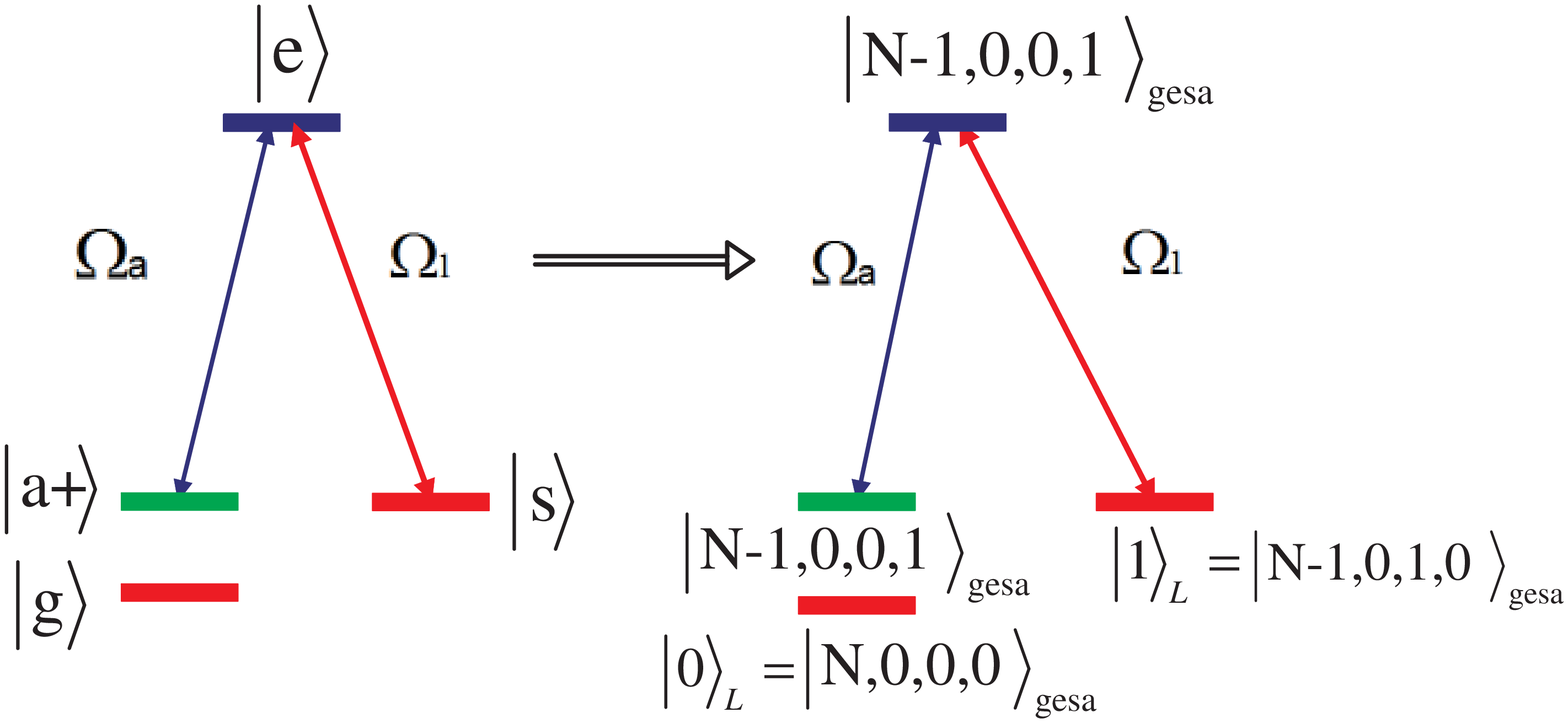}
\caption{\label{Fig:phasegate} Left: Level structure and laser coupling
diagram for a single atom. Right: The equivalent coupling diagram for an ensemble
of atoms in second quantized representation.}
\end{figure}

We first consider how to perform a simple one qubit phase gate:
$|1\rangle_L\rightarrow e^{i\phi_1}|1\rangle_L$. To simplify the model,
we only consider the energy levels we are interested in. The total
pulse sequence and resonant coupling diagram is shown in Fig.~\ref{Fig:phasegate}. The Hamiltonian of of this system,
in the interaction picture and rotating wave approximation, is
\begin{equation}\label{H1}
\begin{split}
H_1(t) = & \sum_i\Omega^i_1(t)\bigl(\sigma_{se}^{(i)}+ \textrm{H.c} \bigr) \\
&+\sum_i\Omega^i_a(t)\bigl(\sigma_{a_+e}^{(i)}+ \textrm{H.c}\bigr),
\end{split}
\end{equation}
where the $\sigma_{jk}^{(i)}=|j_i\rangle\langle k_i|$ are transition operators
from state $|k\rangle$ to $|j\rangle$ for atom $i$, and $\hbar=1$.
Here, the two lasers are assumed to be phase matched,
and their Rabi frequencies are real numbers. For simplicity, assuming the laser beams
illuminate each atom with same intensity, we can rewrite the
Hamiltonian in the second quantized representation:
\begin{equation}\label{H1_2}
\begin{split}
H_1(t) =& \Omega_1(t)\mathcal {\hat{E}}^{\dagger}\mathcal{\hat{S}}
+ \Omega_1^{*}(t)\mathcal {\hat{S}}^{\dagger}\mathcal{\hat{E}} \\
&+ \Omega_a(t)\mathcal {\hat{E}}^{\dagger}\mathcal{\hat{A}}_+
+ \Omega_a^{*}(t)\mathcal {\hat{A}}_+^{\dagger}\mathcal{\hat{E}},
\end{split}
\end{equation}
and set $\Omega_1 = \Omega\sin\theta$ and
$\Omega_a = -\Omega\cos\theta e^{i\varphi}$.
The parameters $\theta$ and $\varphi$ are functions of time, and
the absolute magnitude $\Omega$ is a constant
that must be large enough to satisfy the adiabatic condition.
$\mathcal {\hat{E}},\mathcal {\hat{S}},\mathcal {\hat{A}}_+$ are
bosonic operators defined in last section corresponding to the single
atom states $|e\rangle,|s\rangle,|a_+\rangle $, respectively.

\begin{figure}[!htp1]
\centering\includegraphics[width=65mm,height=80mm]{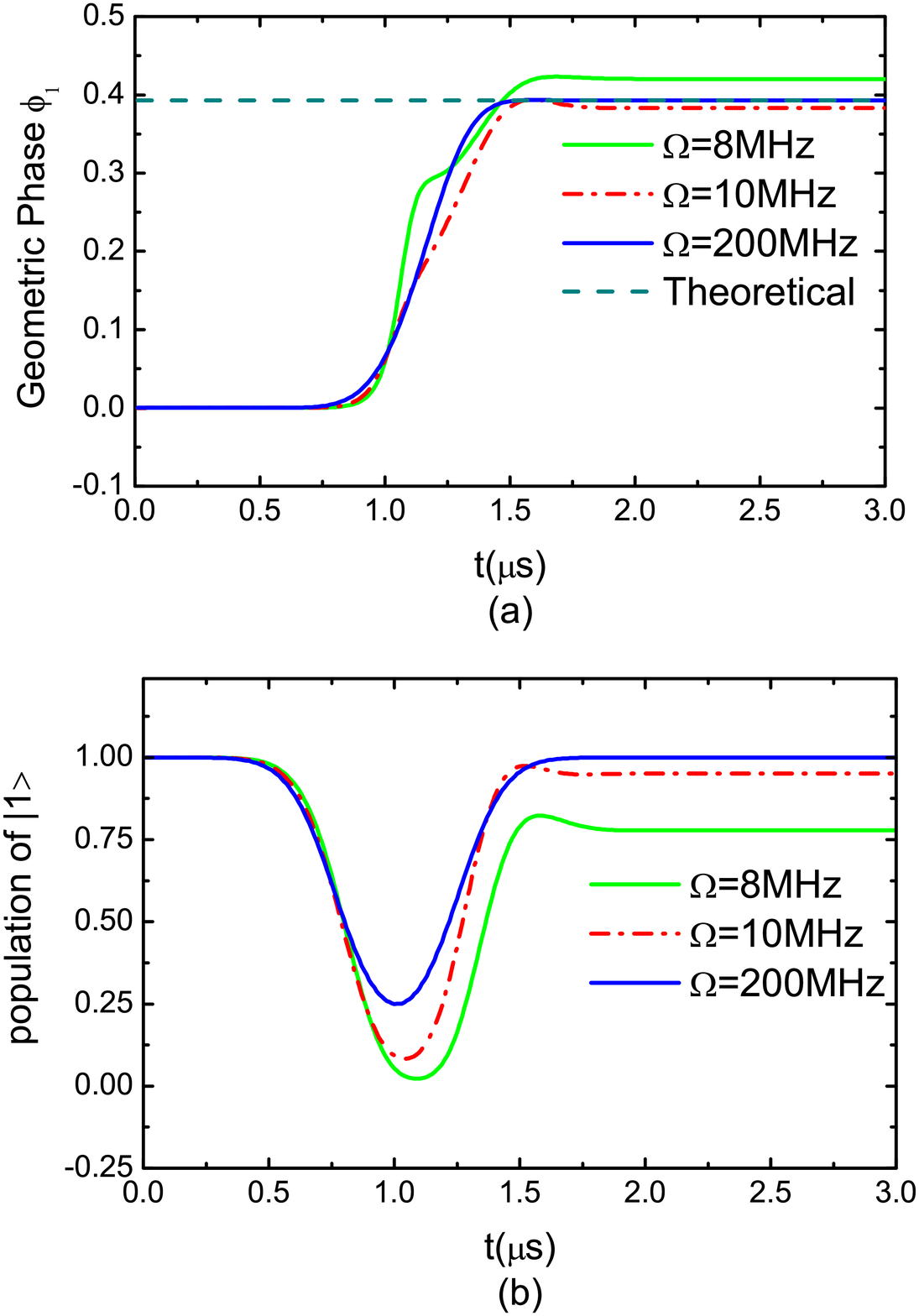}
\caption{\label{Fig:singlephase} (a) The evolution of the relative phase between
state $|1\rangle_L$ and state $|0\rangle_L$ for three different values
of $\Omega$, to realize a $\pi/16$ gate. Note that after the cyclic evolution, we get an
additional pure geometric phase, which fits the theoretical value quite well for $\Omega$
larger than 200MHz. (b) The population change of state $|1\rangle_L$ during the
process. We can see that only if $\Omega$ is large enough to satisfy the adiabatic
condition can we get high accuracy of the gate. Two short laser pulses are
determined by $\theta(t)=\exp((t-1)^2/0.15)$ and $\varphi(t)=\exp((t-1.5917)^2/0.15)$
(the unit of $t$ is $\mu$s).}
\end{figure}

We can regard the system as three coupled harmonic oscillators.
We are only interested in singly excited states; the Hamiltonian discussed above
is closed in that basis, which for simplicity we represent as
$\{|\textbf{e}\rangle = |N-1,1,0,0\rangle_{gesa}$,
$|1\rangle_L = |\textbf{s}\rangle = |N-1,0,1,0\rangle_{gesa}$,
$|\textbf{a}_+\rangle = |N-1,0,0,1\rangle_{gesa}\}$. The
coupling diagram of the ensemble of atoms is
equivalent to a simple three-state coupling diagram for a single atom,
as shown in Fig.~\ref{Fig:phasegate}.

Representing the Hamiltonian in matrix form in this basis, we have
\begin{equation}
H_1(t) = \left[
  \begin{array}{ccc}
    0 & \Omega\sin\theta & -\Omega\cos\theta e^{i\varphi} \\
   \Omega\sin\theta  & 0 & 0 \\
   -\Omega\cos\theta e^{-i\varphi} & 0 & 0 \\
  \end{array}
\right].
\end{equation}
The zero-energy eigenspace of the Hamiltonian is
non-degenerate. The dark state is
\begin{equation}
|D(t)\rangle = \cos\theta|1\rangle_L + \sin\theta
e^{-i\varphi}|\textbf{a}_+\rangle.
\end{equation}
Note that the Hamiltonian is decoupled from the state $|0\rangle_L =
|N,0,0,0\rangle_{grsa}$.

Assume the initial state is
$|\psi(-\infty)\rangle = |D(-\infty)\rangle = |1\rangle_L$ (which means
$\theta(-\infty) = 0$ at the very beginning).  Also, assume that initially
$\varphi = 0$. Now, turn on two laser beams, using
the standard formula to calculate the geometric phase, where the
parameters make a cyclic evolution with the starting and ending point
$\theta = 0, \varphi = 0$:
\begin{equation}
\phi_1 = i\oint \textrm{d}\textbf{R}\langle
D(t)|\nabla_{\textbf{R}}|D(t)\rangle,
\end{equation}
where $\textbf{R}(t) = (\theta(t), \varphi(t))$. We have
\[
\langle D(t)|\nabla_{\textbf{R}}|D(t)\rangle = -i\sin^2(\theta)\hat{\varphi}
\]
and
\[
\phi_1 = \oint\sin^2(\theta)\textrm{d}\varphi .
\]
According to Green's theorem, this geometric phase is exactly the enclosed solid angle:
\[
\phi_1 = \iint_{\partial S}\sin\theta \textrm{d}\theta\textrm{d}\varphi
 = \iint_{\partial S}\textrm{d}\Omega .
\]
A pulse sequence consisting of two stimulated Raman
adiabatic passage (STIRAP) pulses with relatively large
width (to guarantee that the process is adiabatic)
can be  used for this purpose.

A numerical simulation of resonance coupling to realize a $\pi/16$ gate ($\phi_1=\pi/8$) is shown in Fig.~\ref{Fig:singlephase}. Since in the laboratory the Rabi frequency $\Omega$ can be as high as 200 MHz, the adiabatic process could be finished in several microseconds, which is much shorter than the decoherence time of the hyperfine ground states.  We can exchange the population of $|1\rangle_L$ and $|{\bf a}_+\rangle$ back and forth without losing any information, and we can get a geometric phase with high accuracy if $\Omega$ is large enough.

In the previous method of using ensembles of atoms or molecules to realize quantum information processing, a major shortcoming is that it is very hard to uniformly illuminate all the atoms, thus limiting the accuracy of the gate operation. But in our proposal, each atoms interacts individually with $\Omega_1$ and $\Omega_a$, and the geometric phase is independent of the absolute value of $\Omega$. If the spatial distribution of the electric fields of the two laser beams match each other, say $\widetilde{\Omega}_1(\textbf{r})=\widetilde{E}_1(\textbf{r})d_{es}= \widetilde{E}_a(\textbf{r})d_{ea_+}=\widetilde{\Omega}_a(\textbf{r})$ for every point in space $\textbf{r}$, the effect of non-uniform illumination can be naturally eliminated. This is an advantage of adiabatic control.

\subsection{Y-Rotation Gate}

Now we show how to achieve the gate $U = e^{i\phi_2 \sigma_y}$. This gate is much more complicated to realize than the one qubit phase gate. It is difficult to couple the state $|0\rangle_L$ to $|1\rangle_L$, since directly coupling them by applying a laser beam to the ensemble of atoms would result in unwanted higher collective excitations.  In Ref. \cite{MDLukin2001}, a scheme taking advantage of the dipole-dipole interaction of Rydberg states between atoms was used to realize this gate. Here we propose a geometric gate by following three steps:

\begin{enumerate}
  \item Adiabatically pump the ground state $|0\rangle_L$ to a highly excited Rydberg state $|\textbf{r}\rangle$, using the dipole-dipole interaction to assure that only a single collective excitation is achieved.
  \item Adiabatically control the coupling between the single excitation states $|\textbf{r}\rangle$, $|\textbf{s}\rangle$, and $|\textbf{a}_-\rangle$ to geometrically realize the gate.
  \item Reverse step 1 to transfer $|\textbf{r}\rangle$ back to $|0\rangle_L$.
\end{enumerate}

To carry out step 1, consider Rydberg states of a hydrogen atom within a
manifold of fixed principle quantum number $n$
with degeneracy $n^2$. This degeneracy can be removed by applying a
constant electric field $E$ along a certain axis
(linear Stark effect), such as the $z$ axis. For electric
fields below the Ingris-Teller limit, the mixing of adjacent $n$
manifolds can be ignored, and energy levels are split according to
$\Delta E_{nqm} = 3nqea_{B}E/2$ with parabolic and
magnetic quantum numbers $q$ and $m$, respectively. Here, $q$ can take values
$n-1-|m|, n-3-|m|,...,-(n-1-|m|)$, $e$ is the electron charge, and
$a_B$ the Bohr radius. These states have dipole moments of $\mu =
3nqea_B\textbf{e}_z/2$.

Consider two atoms $k$ and $l$ seperated by a distance $\textbf{R}$. The dipole-dipole interaction between them is
\begin{equation}
V_{\textrm{dip}}^{kl}(\textbf{R}) = \frac{1}{4\pi\epsilon_{0}}
\left[\frac{\hat{\mu}_k\cdot\hat{\mu}_l}{|\textbf{R}|^3}-3\frac{(\hat{\mu}_k\cdot
\textbf{R})(\hat{\mu}_l\cdot \textbf{R})}{|\textbf{R}|^5}\right],
\end{equation}
where $\hat{\mu}_k$ and $\hat{\mu}_l$ are dipole moment operators
for atoms $k$ and $l$. Suppose the electric field is sufficiently large that the
energy splitting between two adjacent Stark states is much larger
than the dipole-dipole interaction.

For two atoms initialized in Stark eigenstates, the diagonal terms of
$V_{\textrm{dip}}^{kl}(\textbf{R})$ provide an energy shift, while
the nondiagonal terms couple adjacent $m$ manifolds with each other,
$(m,m)\rightarrow (m\pm1,m\mp1)$. The off-diagonal transition can cause
decoherence, and is suppressed by an appropriate choice of
initial Stark eigenstate \cite{Neutral_atom2}. (For hydrogen, the
state is $|r\rangle_n=|n, q=n-1,m=0\rangle$.) For a fixed distance
$\textbf{R} = \textrm{R}\textbf{e}_z$, the dipole-dipole interaction
of two atoms $k$ and $l$ in the states $|r_k\rangle_{n_1}$ and $|r_l\rangle_{n_2}$ is
\begin{eqnarray*}
u^{kl}_{n_1n_2}(\textrm{R}) &=& {}_{n_1}\langle r_k|
{}_{n_2}\langle r_l | V^{kl}_{\textrm{dip}}(\textbf{R})  |r_k\rangle_{n_1}
|r_l\rangle_{n_2}  \\
&=& - \frac{9 a_B^2 e^2}{R^3 8\pi\epsilon_0}[n_1n_2(n_1-1)(n_2-1)]  .
\end{eqnarray*}
For $n_1$ and $n_2$ sufficiently large, $u(\textrm{R}) \propto n_1^2n_2^2$. For
alkali atoms such as rubidium, the situation is more complicated, but
this kind of analysis still works. In the appendix, we discuss how to pick a state
which has characteristics similar to $|r\rangle_n$ for rubidium.
We will make use of this energy shift later.

\begin{figure}
\centering\includegraphics[width=80mm,height=50mm]{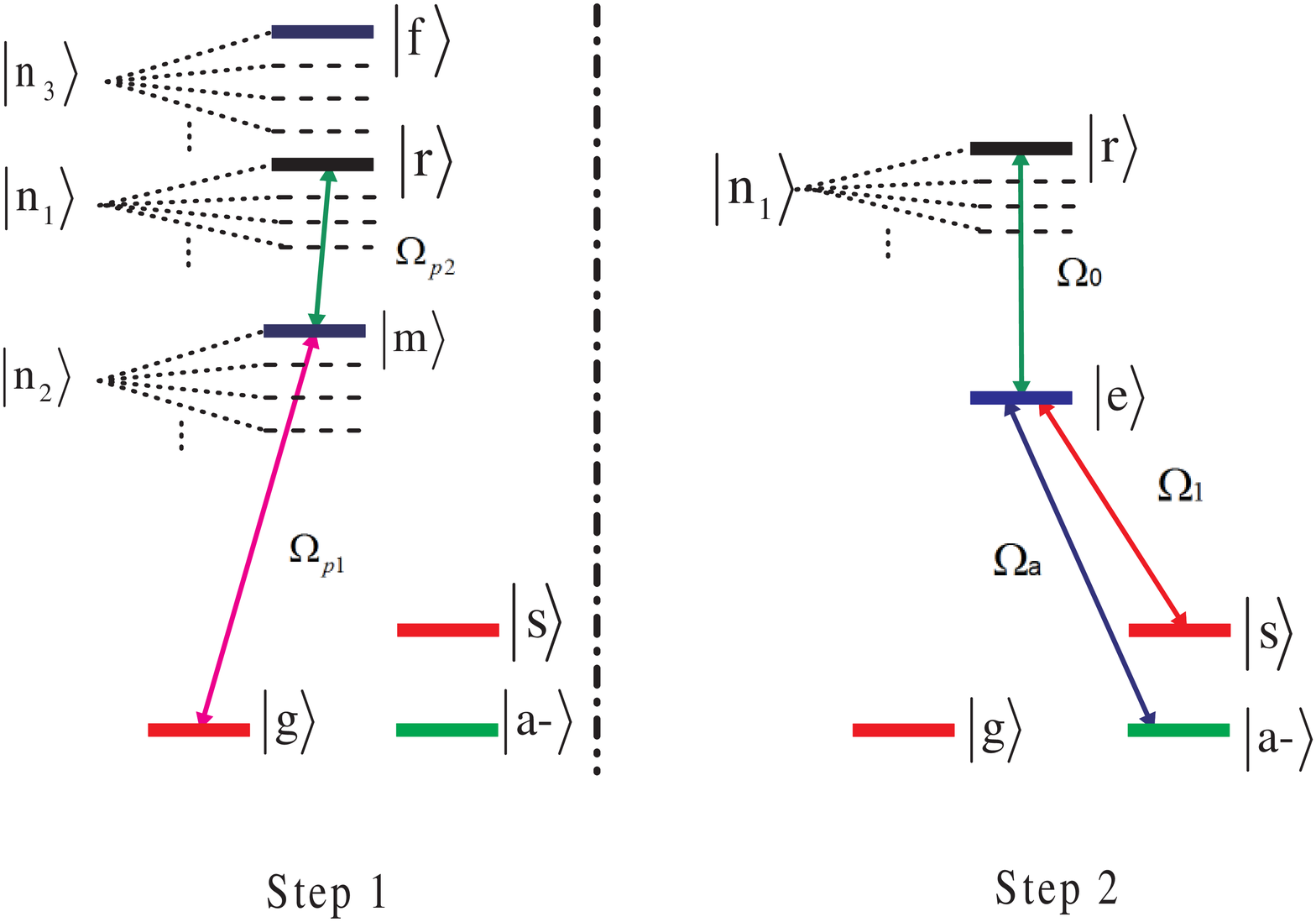}
\caption{\label{Fig:Rygate} The level diagram with the couplings to
realize the $R_y^i(\phi)$ gate. We apply an electric field to the
ensemble of atoms, and pick appropriate Stark eigenstates of rubidium to be the
register states $|r\rangle = |r\rangle_{70,0}$, $|f\rangle=|r\rangle_{72,0}$ and the intermediate Rydberg state $|m\rangle = |r\rangle_{60,0}$ as shown the Appendix.
States $|g\rangle$, $|a_-\rangle$ and $|s\rangle$ are the in the same
manifold of the internal ground state. In the first step, two conjugate laser beams $\Omega_{p1}$ and $\Omega_{p2}$ are used to adiabatically pump a single excitation from $|\textbf{g}\rangle$ to $|\textbf{r}\rangle$. In the second step, three laser beams $\Omega_0$, $\Omega_1$, $\Omega_a$ are used to adiabatically realize the $R_y^i(\phi)$ gate. The excitation from $|g\rangle$ to $|r\rangle$ is then adiabatically reversed.}
\end{figure}

Consider the diagram shown in Fig.~\ref{Fig:Rygate}. For step 1, there should be a single excitation in state $|r\rangle$ after pumping if the initial state is $|0\rangle_L$, and no excitation in state $|r\rangle$ if the initial state is $|1\rangle_L$. To achieve this goal, we introduce two other Rydberg states $|m\rangle$ and $|f\rangle$, and adiabatically transfer the amplitude of $|1\rangle_L$ to $|{\bf f}\rangle$ by coupling $|s\rangle$ and $|f\rangle$ to some intermediate state, say $|e\rangle$. Please note that this pumping process can also be realized adiabatically.

The Hamiltonian of the cloud of atoms as described in the left side of Fig.~\ref{Fig:Rygate} in the
rotating wave approximation is:
\begin{equation}
\begin{split}
H_{p}(t) =  \sum_{i} & (\Omega_{p1}(t)\sigma_{gm}^{(i)}+\textrm{H.c})\\
+\sum_{i}&(\Omega_{p2}(t)\sigma_{mr}^{(i)}+\textrm{H.c})\\
 + \sum_{k>l} & \biggl( u^{kl}_{mm} (\textbf{R}_{kl})|m_k\rangle|m_l\rangle\langle m_k|\langle m_l|\\
&+u^{kl}_{rr}(\textbf{R}_{kl})|r_k\rangle|r_l\rangle\langle r_k|\langle r_l|)\\
&+u^{kl}_{ff}(\textbf{R}_{kl})|f_k\rangle|f_l\rangle\langle f_k|\langle f_l|\\
&+u^{kl}_{rm}(\textbf{R}_{kl})|r_k\rangle|m_l\rangle\langle r_k|\langle m_l |\\
&+u^{kl}_{fm}(\textbf{R}_{kl})|f_k\rangle|m_l\rangle\langle f_k|\langle m_l|\\
&+u^{kl}_{fr}(\textbf{R}_{kl})|f_k\rangle|r_l\rangle\langle f_k|\langle r_l| \biggr) .
\end{split}
\end{equation}
The two-body interaction term here is due to the energy shift of the states. Just as in the case of the phase gate, we represent the Hamiltonian in second quantized form. Again, assuming the laser fields are uniformly coupled to each atom, and using the standard second quantization procedure to deal with the two-body interaction, we apply the minimum energy shift for all pairs of atoms (as the worst case) and obtain
\begin{equation}
\begin{split}
H^{\textrm{eff}}_p(t)\approx & \Omega_{p1}(t)\sqrt{N}(\mathcal{\hat{M}}^{\dagger}
 + \mathcal{\hat{M}})\\
 &+ \Omega_{p2}(t)(\mathcal{\hat{R}}\mathcal{\hat{M}}^{\dagger}
 + \mathcal {\hat{R}}^{\dagger}\mathcal {\hat{M}}) \\
&+ \overline{u}_{mm}\mathcal {\hat{M}^\dagger}\mathcal{\hat{M}^\dagger}
  \mathcal{\hat{M}}\mathcal {\hat{M}}
  + \overline{u}_{rr}\mathcal{\hat{R}^\dagger}\mathcal{\hat{R}^\dagger}
  \mathcal{\hat{R}}\mathcal{\hat{R}}\\
& + \overline{u}_{ff}\mathcal{\hat{F}^\dagger}\mathcal{\hat{F}^\dagger}
 \mathcal{\hat{F}}\mathcal{\hat{F}}
 + \overline{u}_{rm}\mathcal{\hat{R}^\dagger}\mathcal{\hat{R}}
 \mathcal {\hat{M}^\dagger}\mathcal{\hat{M}}\\
& + \overline{u}_{fm}\mathcal{\hat{F}^\dagger}\mathcal{\hat{F}}
 \mathcal{\hat{M}^\dagger}\mathcal{\hat{M}}
 + \overline{u}_{fr}\mathcal{\hat{F}^\dagger}\mathcal{\hat{F}}
 \mathcal {\hat{R}^\dagger}\mathcal{\hat{R}},
\end{split}
\end{equation}
where $\overline{u}_{rr}$, $\overline{u}_{mm}$, $\overline{u}_{ff}$, $\overline{u}_{rm}$, $\overline{u}_{fm}$ and $\overline{u}_{fr}$ are chosen to be
the minimum energy shifts from the dipole-dipole interaction.

\begin{figure}[!h]
\centering\includegraphics[width=80mm,height=60mm]{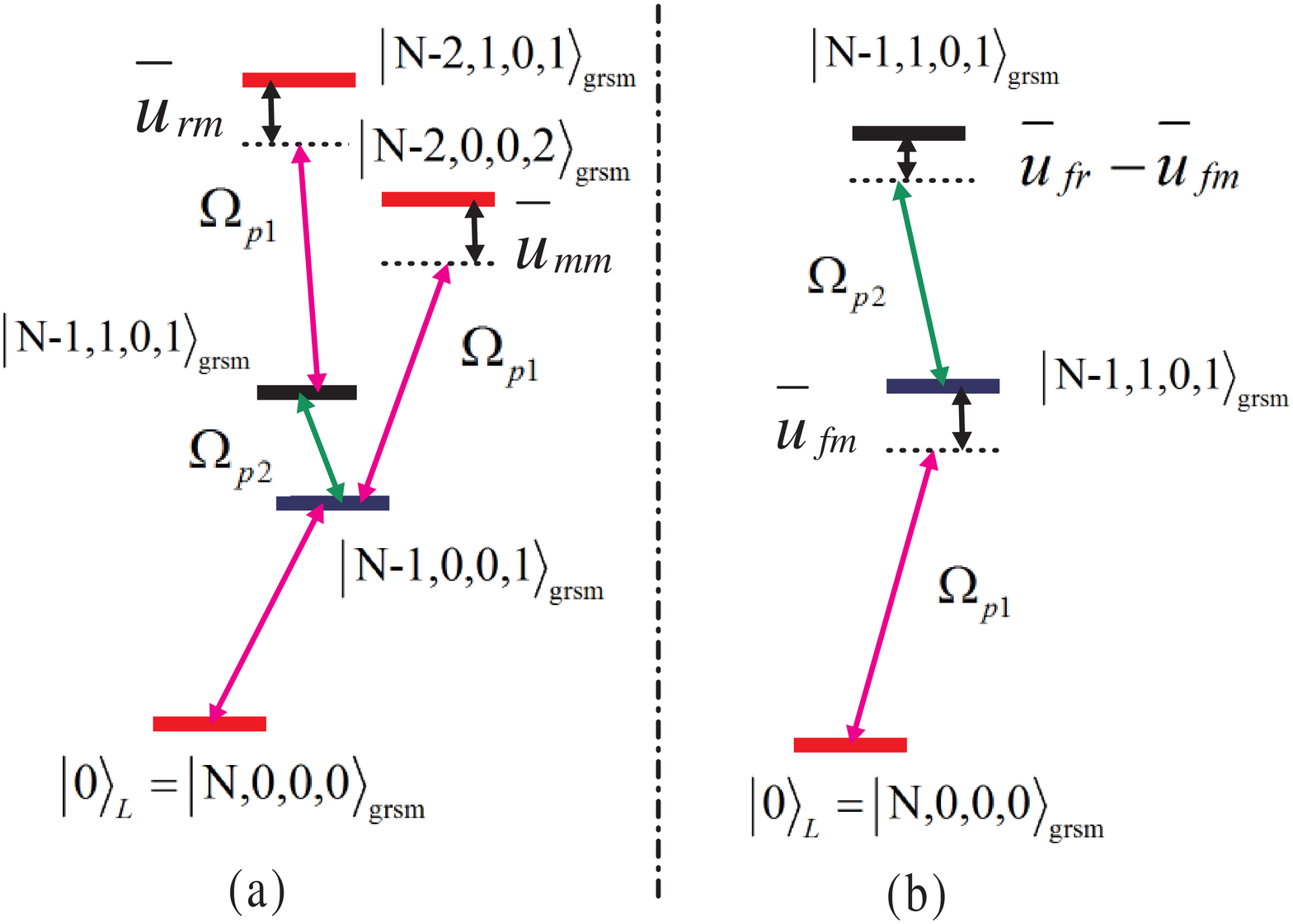}
\caption{\label{Fig:eq-ry}The equivalent coupling diagram of the
ensembles of atoms in cases of collective excitation for the case (a) system is initially in $|0\rangle_L$ and (b) system is initially in $|1\rangle_L$.}
\end{figure}

In the case of low excitation, in the Bogoliubov approximation we can
regard the operators $\mathcal {G}, \mathcal {G}^{\dagger}$ on state
$|g\rangle$ as the $C$-number $\sqrt{N}$, where $N$ is the
total number of the atoms in the cloud. When the system is initially in
$|0\rangle_L$ (and hence $|f\rangle$ is not excited), the
equivalent state coupling diagram for this pumping process is
shown in Fig.~\ref{Fig:eq-ry}(a). Unlike the phase gate, we see that
this Hamiltonian is no longer closed in a small
subspace of states because $\Omega_{p1}$ will continuously pump
atoms of the ensemble from the ground state to the excited states.
However, because of the large energy shift, the rate to excite
$|N-2,1,0,1\rangle_{grsm}$ and $|N-2,1,1,0\rangle_{grsm}$ is strongly suppressed.
Physically, this means there cannot be an atom excited to the Rydberg
state $|r\rangle$ while another atom be excited to state
$|i\rangle$ or $|m\rangle$. Only a single excitation to the register state
can be achieved ($|0\rangle_L\rightarrow|\textbf{r}\rangle$)
during this adiabatic process.

\begin{figure}
\centering\includegraphics[width=70mm,height=60mm]{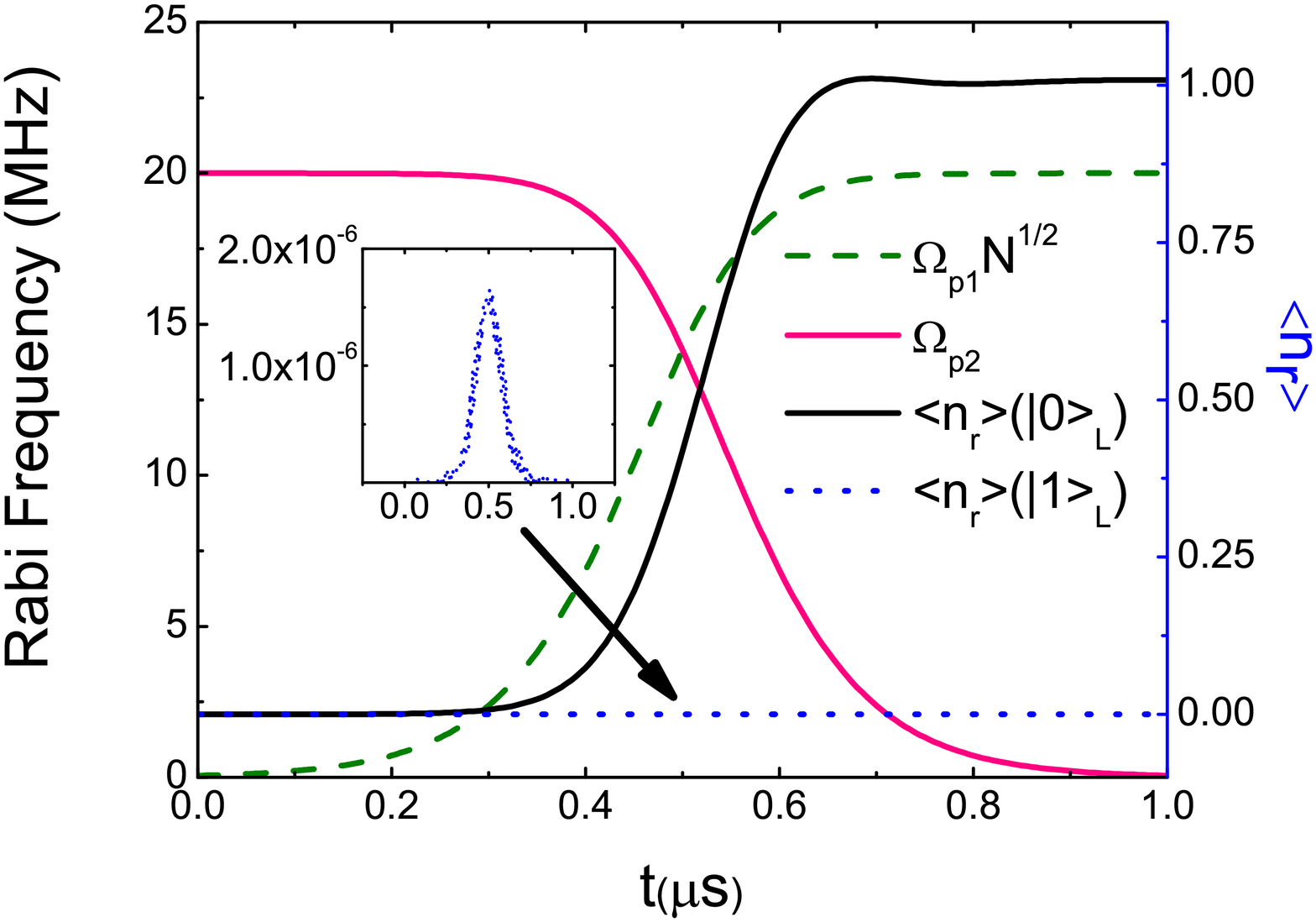}
\caption{\label{Fig:pump} Adiabatic Pumping process. Note that at the end there is
only a single excitation in the register state $|r\rangle$ when initially in $|0\rangle_L$, and no excitation when initially in $|1\rangle_L$. Here, we set $\overline{\mu}_{rm} = 400\mbox{MHz}$ and $\overline{\mu}_{mm} 300\mbox{MHz}$. 
Since we take the minimum energy shift as our simulation parameter,
the actual performance should be better (i.e., the process could
be finished in a shorter period of time).}
\end{figure}

If the system is initially in the state $|1\rangle_L$
(and hence there is a single excitation in $|f\rangle$), both
$|r\rangle$ and $|m\rangle$ cannot be excited due to the
energy shift caused by the dipole-dipole interaction, as
shown in Fig.~\ref{Fig:eq-ry}(b). A numerical simulation
of the pumping process for the minimum energy shifts is
shown in Fig.~\ref{Fig:pump}.

\begin{figure*}
\centering\includegraphics[width=130mm,height=100mm]{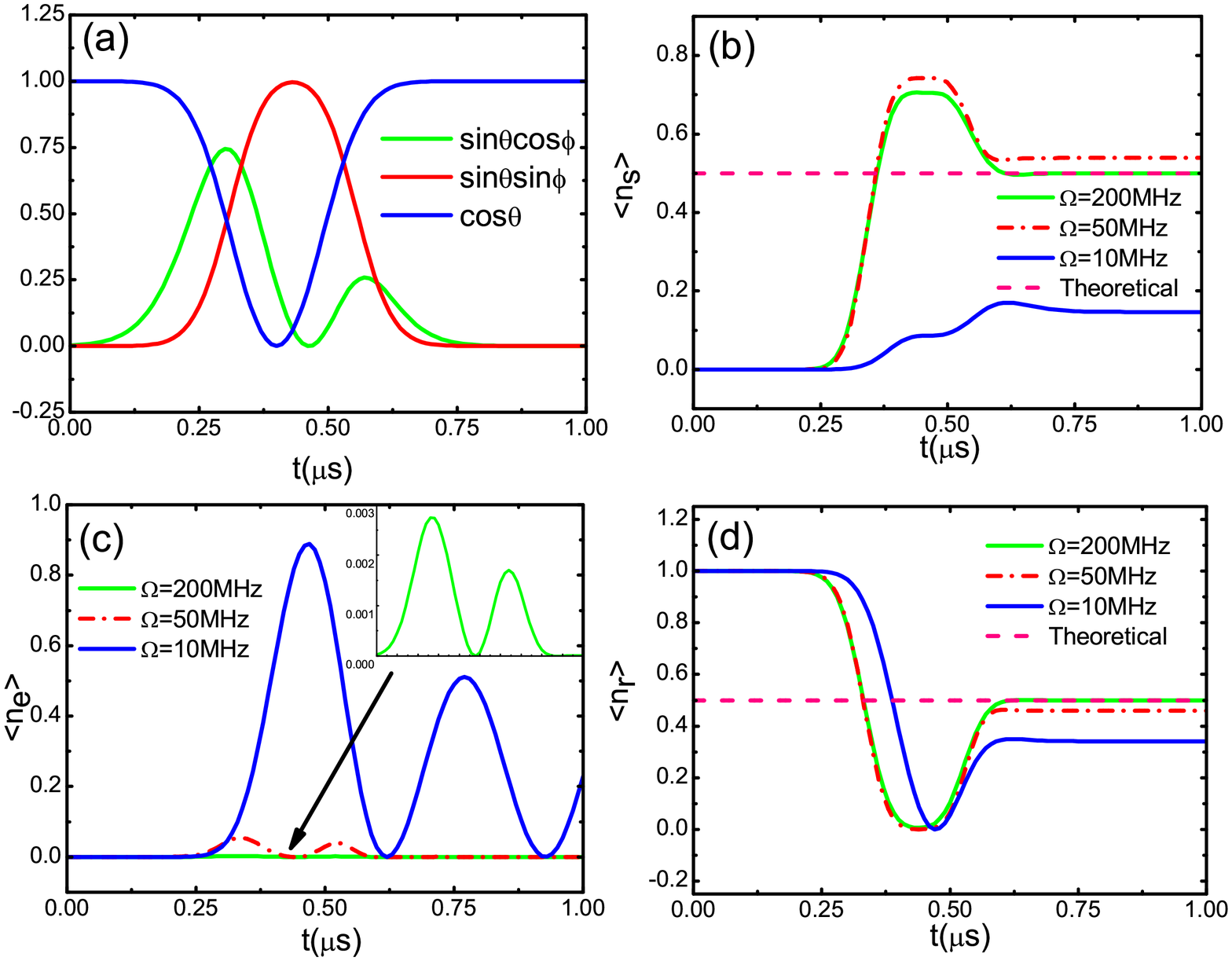}
\caption{\label{ygate.plot} (a): The pulse shape of the control lasers of $\Omega_0$, $\Omega_1$ and $\Omega_a$ respectively. (b)(c)(d): The evolution of the populations of $|s\rangle$, $|e\rangle$, $|r\rangle$. We see that when $\Omega$ is larger than 200MHz, the adiabatic condition is satisfied and the gate works with very high accuracy (close to the theoretical value) and the population of the short-lived state $|e\rangle$ tends to zero.}
\end{figure*}

After the pumping process, and the adiabatic transfer of
the population of $|{\bf f}\rangle$ back to $|1\rangle_L$,
we can then realize the $R_y^i(\phi)$ gate on the ensemble
of atoms by applying three controlling laser beams. The coupling
diagram is shown in the right side of Fig.~\ref{Fig:Rygate}.
The Hamiltonian of step 2 can be obtained directly in
second-quantized form:
\begin{equation}
\begin{split}
H_2(t) =&\Omega_0(t)(\mathcal {\hat{E}}\mathcal
{\hat{R}}^{\dagger} + \mathcal {\hat{E}}^{\dagger}\mathcal
{\hat{R}})+\Omega_1(t)(\mathcal{\hat{E}}\mathcal {\hat{S}}^{\dagger} +
\mathcal {\hat{E}}^{\dagger}\mathcal {\hat{S}})\\
&+ \Omega_a(t)(\mathcal
{\hat{E}}\mathcal{\hat{A}}^{\dagger}_{-}+\mathcal{\hat{E}}^{\dagger}\mathcal {\hat{A_-}}).\\
\end{split}
\end{equation}
We can directly apply the technique of geometric transformations to realize this gate,
as introduced in Ref. \cite{L-M-Duan_Science_2001} for four-level
systems. Here, we choose $\Omega_0=
\Omega\sin\theta\cos\varphi$, $\Omega_1 =
\Omega\sin\theta\sin\varphi$, $\Omega_a = \Omega\cos\theta$.  This time, the Hamiltonian is closed in
the basis $\{|\textbf{e}\rangle, |0\rangle_L = |\textbf{r}\rangle, |1\rangle_L = |\textbf{s}\rangle,
|\textbf{a}_-\rangle\}$:
\begin{equation}
H_2(t)=\Omega\left[
             \begin{array}{cccc}
               0 & \sin\theta\cos\varphi & \sin\theta\sin\varphi & \cos\theta\\
               \sin\theta\cos\varphi & 0 & 0 & 0 \\
               \sin\theta\sin\varphi & 0 & 0 & 0 \\
               \cos\theta & 0 & 0 & 0 \\
             \end{array}
           \right].
\end{equation}
The eigenspace corresponding to the zero-energy eigenvalue (dark space) is spanned by
basis vectors $\{|D_1\rangle$, and $|D_2\rangle\}$, where
\[
|D_1\rangle = \sin\varphi|0\rangle_L - \cos\varphi|1\rangle_L ,
\]
\[
|D_2\rangle = \cos\theta(\cos\varphi|0\rangle_L + \sin\varphi|1\rangle_L)
 - \sin\theta|\textbf{a}_-\rangle .
\]
We can use the formula for the degenerate subspace under the cyclic
evolution of $\theta$, $\varphi$. Suppose $|\psi(t)\rangle =
C_1(t)|D_1\rangle + C_2(t)|D_2\rangle$. We have the equation
\begin{equation}
\left[
  \begin{array}{c}
    \dot{C_1} \\
    \dot{C_2 }\\
  \end{array}
\right] =\left[
            \begin{array}{cc}
              D_{11}(t) & D_{12}(t) \\
              D_{21}(t) & D_{22}(t) \\
            \end{array}
          \right]\left[
                   \begin{array}{c}
                     C_1 \\
                     C_2 \\
                   \end{array}
                 \right],
\end{equation}
where $D_{ij}(t) = \langle D_i|\dot{D}_j\rangle$. Let's define
\begin{equation}
\begin{split}
 d\emph{A}_2(t)&=\left[
                           \begin{array}{cc}
                            D_{11}(t) & D_{12}(t) \\
                            D_{21}(t) & D_{22}(t) \\
                           \end{array}
                         \right] dt \\
                         &= \left[
                                     \begin{array}{cc}
                                       0 & -\cos\theta\dot{\varphi} \\
                                       \cos\theta\dot{\varphi} & 0 \\
                                     \end{array}
                                   \right] dt ,
\end{split}
\end{equation}
and also the unitary matrix
\begin{equation}
\mathcal {U} = \mathcal{P}\exp(i\oint_{\textbf{C}} d\emph{A}_2(t)).
\end{equation} After the cyclic evolution,
\begin{equation}
\begin{split}
|D_1(T)\rangle &= \mathcal {U}^{11}|D_1(0)\rangle + \mathcal
{U}^{21}|D_2(0)\rangle,\\
|D_2(T)\rangle &= \mathcal {U}^{12}|D_1(0)\rangle + \mathcal
{U}^{22}|D_2(0)\rangle.
\end{split}
\end{equation}
At the beginning, if we set $\theta(0) = 0$ and $\varphi(0) =
\frac{\pi}{2}$, we will have $|D_1(0)\rangle = |0\rangle_L$ and $|D_2(0)
= |1\rangle_L$. We can use the Dyson expansion to get
the unitary matrix $\mathcal {U}$. This gives us:
\begin{equation}
\begin{split}
\mathcal {U}^{11} &= 1 - \frac{\phi_2^2}{2} + \frac{\phi_2^4}{24} - \frac{\phi_2^6}{720}... = \cos\phi_2,\\
\mathcal {U}^{12} &= \phi_2 - \frac{\phi_2^3}{6} + \frac{\phi_2^5}{120}... = \sin\phi_2,\\
\mathcal {U}^{21} &= -\phi_2 + \frac{\phi_2^3}{6} - \frac{\phi_2^5}{120}... = -\sin\phi_2, \\
\mathcal {U}^{22} &= 1 - \frac{\phi_2^2}{2} + \frac{\phi_2^4}{24} - \frac{\phi_2^6}{720}... = \cos\phi_2 ,
\end{split}
\end{equation}
where
\[
\phi_2 = \oint_{\textbf{C}}\cos\theta\textrm{d}\varphi
\]
is a pure geometric phase. Thus we get the unitary transformation we want. Step 3 is simply the reverse of step 1, to return us to our original basis.

A simulation to realize $\phi_2=\pi/4$ was done for a certain pulse shape of the coupling laser beams, where the qubit was initially prepared in the state $|0\rangle_L$, as  shown in Fig.~\ref{ygate.plot}(a). We show the evolution of the populations of states $|s\rangle$, $|e\rangle$ and $|r\rangle$ in Fig.~\ref{ygate.plot}(b),(c), and (d), respectively. If $\Omega$ is large enough ($\geq 200$ MHz, which is practical
in current experiments), we could implement the gate operation in 1$\mu$s with extremely high accuracy. Thus, the total procedure takes less than 3$\mu$s to complete, while the lifetime of $|\textbf{r}\rangle$ is approximately equal to the lifetime of $|r\rangle$ \cite{Lukin_EIT_Review}, which is estimated to be 300-400$\mu$s in a cryogenic environment \cite{cryogenic}.

We simplified the calculation and simulation by assuming that all atoms are uniformly illuminated in the laser beam. However, just like the argument made in the previous section, the geometric phase is independent for large $\Omega$, if the spatial distribution of the electric fields of the three laser beams match each other at each point in space. Thus, we can in principal eliminate the error introduced by nonuniform illumination.

\subsection{Two Qubit Controlled Phase Gate}

\begin{figure}
\centering\includegraphics[width=80mm,height=70mm]{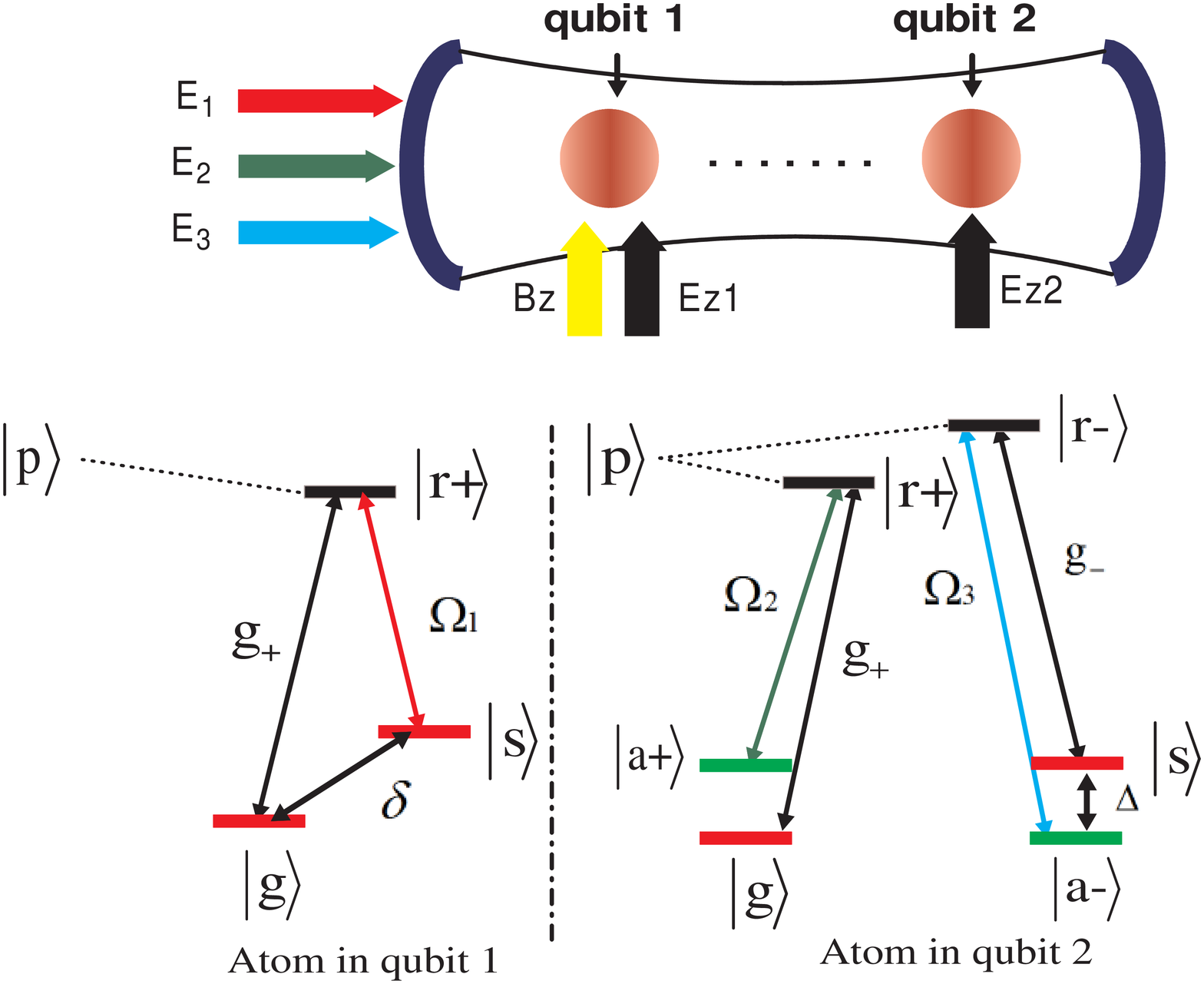}
\caption{\label{c-phase-illustration} The schematic setup to realize the conditional phase gate, and
the relevant atomic levels of the two clouds of atoms.}
\end{figure}

Now, we are ready to construct the controlled phase gate. Recently, an adiabatic SWAP operation between states of two clouds of atoms in a cavity QED system was experimentally achieved \cite{Swap}. Here, we extend this method to geometrically realize the controlled phase gate, which together with the earlier one-qubit gates gives a universal set of quantum gates. The basic idea is to couple two qubits (that is, clouds of atoms) by virtually emitting and absorbing a cavity photon; and after an adiabatic evolution, the second qubit obtains an extra relative phase when the first qubit is in $|0\rangle_L$. The coupling diagram of the scheme is shown
in Fig.~\ref{c-phase-illustration}.

Suppose the length of the cavity is about $200\mu$m, which means, it could contain about 10 qubits. Consider two different clouds of atoms in the cavity. For a single atom in each cloud, pick a manifold of states of $p$. When an electric field is applied to these two cloud of atoms, the manifold splits into $|r_+\rangle$ and $|r_-\rangle$. For atom cloud 1, we set a constant electrical field $E_{z1}$ and a constant magnetic field $B_{z1}$ in the $z$ direction. The Stark Effect splits the manifold $p$, and the Zeeman effect changes the energy of $|g\rangle$ and $|s\rangle$.

Assume the energy difference between $|g\rangle$ and $|s\rangle$ to be $\delta$. Set the appropriate non-zero value of $E_{z1}$ and $B_{z1}$ so that $|r_+\rangle$ is resonantly coupled to $|g\rangle$ and $|s\rangle$ by a cavity mode $a$ and a control laser $\Omega_1$ respectively.
For atom cloud 2, $E_{z2}$ is set so that $|r_+\rangle$ is resonantly coupled to $|g\rangle$ and $|s\rangle$ by the same cavity mode $a$ and control laser $\Omega_2$, and $|r_-\rangle$
is resonantly coupled to $|s\rangle$ and $|a_-\rangle$ by the cavity mode and a control laser $\Omega_3$. Note that for $\textrm{Rb}^{87}$, the hyperfine structure energy split of the
ground state manifold is $\Delta/2\pi$=6.835GHz, and the energy difference of $|r_+\rangle$ and $|r_-\rangle$ should set to be equal to $\Delta$ for atom cloud 2 for the purpose of resonant
coupling, while the frequencies of $\Omega_1$, $\Omega_2$ and $\Omega_3$ should be $\omega_{cav}-\delta$, $\omega_{cav}-\Delta$ and $\omega_{cav}+\Delta$, where $\omega_{cav}$ is the frequency of cavity mode.

Since the cavity mode is inhomogeneously distributed, the thermal motion (even at extremely low temperature) of atoms, causes the coupling rate between atoms and cavity photons to vary greatly from one to another (by roughly a factor of 2) and thus makes the system difficult to control accurately. To overcome these difficulties, instead of directly applying the laser beam to the clouds of atoms, we use the idea of external laser driving control \cite{FORT} as shown in schematic setup in Fig. 9. Three classical laser fields $E_1(t)$, $E_2(t)$ and $E_3(t)$ are incident on one mirror of the cavity to drive the transition $|r_+\rangle\rightarrow|s\rangle$ for cloud 1, and $|r_+\rangle\rightarrow|a_+\rangle$ and $|r_-\rangle\rightarrow|a_-\rangle$
for cloud 2 through another cavity mode $a^\prime$. We assume for simplicity that $a$ and $a^\prime$ have the same spatial mode structure $\chi(\textbf{r})$ with the same frequency $\omega_{cav}$, so they can be different only in polarization.

The driving fields $E_1(t)$, $E_2(t)$ and $E_3(t)$ are resonant with frequencies $\omega_{cav}-\delta$, $\omega_{cav}-\Delta$ and $\omega_{cav}+\Delta$.  So they are far-off-resonant to the cavity mode with large detuning. $E_1(t)$, $E_2(t)$ and $E_3(t)$ control the time evolution of the Rabi frequencies $\Omega_1$, $\Omega_2$ and $\Omega_3$. To see this, consider the input-output equation for the cavity mode $a^\prime$ \cite{Quantum Optics}:
\begin{equation}\label{input-output}
\frac{da^\prime(t)}{dt}=-i[a(t), H_{sys}]-\frac{\kappa}{2}a^\prime(t)+\sqrt{\kappa}a_{\textrm{in}}^\prime(t),
\end{equation}
where $a_{\textrm{in}}^\prime(t)$ is the field operator for the input driving pulse coupling to the mode $a^\prime$, with
\[
\langle a_{\textrm{in}}^\prime(t)\rangle = E_1(t)+E_2(t)+E_3(t) ,
\]
and
\[
[a_{\textrm{in}}^\prime(t) ,a_{\textrm{in}}^{\prime\dagger}(t^\prime)]=\delta(t-t^\prime) .
\]
Since $a^\prime$ is driven by strong classical pulses, we can treat the mode as classical and thus assume that the interaction between $a^\prime$ and the atoms will not change the state of the cavity mode $a^\prime$. Eq.~(\ref{input-output}) therefore can be modified to be
\begin{equation}
\begin{split}
\frac{da^\prime(t)}{dt} = & -i\omega_{cav}a^\prime(t)-\frac{\kappa}{2}a^\prime(t) \\
& + \sqrt{\kappa}(E_1(t)+E_2(t)+E_3(t)).
\end{split}
\end{equation}
First, we suppose there is only one classical pulse $E_1(t)=\langle  a_{\textrm{in}}^\prime(t)\rangle =\varepsilon_1(t)e^{-i(\omega_{cav}-\delta)t}$, where $\varepsilon_1(t)$ is the slowly varying amplitude of $E_1(t)$. Then, we can write the mean value of $a^\prime$ as
\[
\langle a^\prime(t)\rangle = \alpha(t)e^{-i(\omega_{cav}-\delta)t} ,
\]
which has the solution
\begin{equation}
\begin{split}
\alpha(t)&=\int_{0}^{t}\varepsilon_1(\tau)\exp(-i\delta-\kappa/2)(t-\tau)d\tau \\
         &\approx\frac{\varepsilon_1(t)-e^{(-i\delta-\kappa/2)t}\varepsilon_1(0)}{i\delta+\kappa/2}
\end{split}
\end{equation}
when the characteristic time $T_1$ of $\varepsilon_1(t)$ satisfies $\delta T_1\gg1$.

Assume that $\varepsilon_1(t)$ gradually increases from zero with $\varepsilon_1(0)=0$. Thus we can conclude $\alpha(t)\propto \varepsilon_1(t)$. By choosing an appropriate phase of $\varepsilon_1(t)$, we set $\alpha(t)$ to be real. If we input $E_2(t)$ or $E_3(t)$, the solution will have a similar form. Now, consider the case when three classical pulses are incident on the cavity mirror.  Eq.~(\ref{input-output}) is a linear equation, so, when the three input pulse have different resonant frequencies, the solution should be a superposition of the solutions for three single pulses. Thus, $\langle a^\prime(t)\rangle$ could be represented as
\begin{eqnarray*}
\langle a^\prime(t)\rangle &=& \alpha_1(t)e^{-i(\omega_{cav}-\delta)t}
 + \alpha_2(t)e^{-i(\omega_{cav}-\Delta)t} \\
&& + \alpha_3(t)e^{-i(\omega_{cav}+\Delta)t} .
\end{eqnarray*}
This gives Rabi frequencies $\Omega_i(\textbf{r},t)=d_i\alpha_i(t)\chi(\textbf{r})$ for $i=1,2,3$, where $d_i$ is a coefficient mainly determined by the dipole moment for the corresponding transition.

The frequencies of the three components of $a^\prime$ differ significantly, so we can regard them as three separate pulses. In addition, the energy structure of atoms in cloud 1 is different from those in cloud 2, so $\Omega_1$ will not interact with the atoms in cloud 2. Similarly, $\Omega_2$ and $\Omega_3$ will not interact with atoms in cloud 1. All other clouds in the cavity are all far off resonant from both the cavity mode and the control lasers, and thus can be safely excluded by our gate operation.
Now, we can represent the Hamiltonian for the situation of
of Fig.~\ref{c-phase-illustration} in the interaction picture:
\begin{equation}
\begin{split}
H_3(t) =& \sum_{j_1}\Omega_1(\textbf{r}_{j_1},t)\sigma_{sr_+}^{(j_1)}+\sum_{j_1}g_+(\textbf{r}_{j_1})a^\dagger\sigma_{gr_+}^{(j_1)}\\
  &+ \sum_{j_2}(\Omega_2(\textbf{r}_{j_2}, t)\sigma_{a_+r_+}^{(j_2)} +
  \Omega_3(\textbf{r}_{j_2},t)\sigma_{a_-r_-}^{(j_2)})\\
  &+ \sum_{j_2}(g_+(\textbf{r}_{j_2})a^\dagger\sigma_{gr_+}^{(j_2)}
  +g_-(\textbf{r}_{j_2})a^\dagger\sigma_{sr_-}^{(j_2)})\\
  &+\textrm{H.c}.,
\end{split}
\end{equation}
where $a$ and $a^\dagger$ are annihilation and creation operators of
 cavity mode $a$, and $g_+(\textbf{r})$ and $g_-(\textbf{r})$ are coupling rates of cavity mode $a$ to the transition $|g\rangle\rightarrow|r_+\rangle$ in cloud 1 and $|s\rangle\rightarrow|r_-\rangle$ in cloud 2. We can represent $g_+(\textbf{r})(g_-(\textbf{r}))=\widetilde{g}_+\chi(\textbf{r})(\widetilde{g}_-\chi(\textbf{r}))$,
 where $\widetilde{g}_+$ and $\widetilde{g}_-$ are constants. So we can rewrite the Hamiltonian as:
\begin{equation}
\begin{split}
H_3(t) &= \sum_{j_1}\chi(\textbf{r}_{j_1})(d_1\alpha_1(t)\sigma_{sr_+}^{(j_1)}+\widetilde{g}_+a^\dagger\sigma_{gr_+}^{(j_1)})\\
  &+ \sum_{j_2}\chi(\textbf{r}_{j_2})(d_2\alpha_2(t)\sigma_{a_+r_+}^{(j_2)} + d_3\alpha_3(t)\sigma_{a_-r_-}^{(j_2)}\\
  &+
  \widetilde{g}_+a^\dagger\sigma_{gr_+}^{(j_2)}
  +\widetilde{g}_-a^\dagger\sigma_{sr_-}^{(j_2)})+\textrm{H.c}.,
\end{split}
\end{equation}

We will see soon that the dark state of such a Hamiltonian should be independent of $\chi(\textbf{r})$, and the adiabatic evolution of the system is determined only by the minimum value of $\chi(\textbf{r}_j)$ for all $j$, say $\chi_m$. So, the Hamiltonian can effectively be represented as:
\begin{equation}
\begin{split}
H_3^{\textrm{eff}}(t) &= \sum_{j_1}(\Omega_1(t)\sigma_{sr_+}^{(j_1)}+g_+a^\dagger\sigma_{gr_+}^{(j_1)})\\
  &+ \sum_{j_2}(\Omega_2(t)\sigma_{a_+r_+}^{(j_2)} +
  \Omega_3(t)\sigma_{a_-r_-}^{(j_2)})\\
  &+ \sum_{j_2}(g_+a^\dagger\sigma_{gr_+}^{(j_2)}
  +g_-a^\dagger\sigma_{sr_-}^{(j_2)})+\textrm{H.c}.,
\end{split}
\end{equation}
where $\Omega_i(t)=d_i\alpha_i(t)\chi_m$ for $i=1,2,3$ and $g_+(g_-)=\widetilde{g}_+\chi_m(\widetilde{g}_-\chi_m)$.  As in the previous cases, we transform the Hamiltonian to second
quantized representation. For two clouds of atoms in the Bogoliubov approximation, we get the new Hamiltonian
\begin{equation}
\begin{split}
H^{\textrm{eff}}_3(t) &\approx \Omega_1(t)(\mathcal {\hat{S}^{\dagger}})_1(\mathcal {\hat{R}_{+}})_1
+\Omega_2(t)(\mathcal{\hat{A}_{+}^{\dagger}})_2(\mathcal{\hat{R}_{+}})_2\\
&+\Omega_3(t)(\mathcal{\hat{A}_{-}^{\dagger}})_2(\mathcal{\hat{R}_{-}})_2+g_+\sqrt{N}a^{\dagger}(\mathcal{\hat{R}}_{+})_1\\
&+g_+\sqrt{N}a^{\dagger}(\mathcal{\hat{R}}_{+})_2+g_-a^{\dagger}(\mathcal {\hat{S}^{\dagger}})_2(\mathcal{\hat{R}}_{-})_2+\textrm{H.c}.,
\end{split}
\end{equation}
where $\mathcal {\hat{S}}$, $\mathcal {\hat{R}_{-}}$, $\mathcal {\hat{R}_{+}}$, $\mathcal {\hat{A}_{-}}$, and $\mathcal {\hat{A}_{+}}$ are the bosonic annihilation operators of the single atom states $|s\rangle$, $|r_-\rangle$, $|r_+\rangle$, $|a_-\rangle$, and $|a_+\rangle$, respectively. We set
\begin{equation*}
\begin{split}
\Omega_1(t) = &\Omega_1\sin\theta e^{-i\varphi_1} ,\\
\Omega_2(t) = &\Omega_2\cos\theta e^{-i\varphi_2} ,\\
\Omega_3(t) = &\Omega_3\cos\theta e^{-i\varphi_3} .
\end{split}
\end{equation*}
We also define $g_1=g_+\sqrt{N}$, $g_2=g_+\sqrt{N}$ and $g_3=g_-$ for convenience.

As in the previous sections, for each cloud of atoms we have a set of basis states:
\begin{align*}
|0\rangle_L &=|N,0,0,0,0,0\rangle_{gr_{+}r_{-}a_{+}a_{-}s},\\
|1\rangle_L &=|N-1,0,0,0,0,1\rangle_{gr_{+}r_{-}a_{+}a_{-}s},\\
|\textbf{r}_+\rangle &=|N-1,1,0,0,0,0\rangle_{gr_{+}r_{-}a_{+}a_{-}s},\\
|\textbf{r}_-\rangle &=|N-1,0,1,0,0,0\rangle_{gr_{+}r_{-}a_{+}a_{-}s},\\
|\textbf{a}_+\rangle &=|N-1,0,0,1,0,0\rangle_{gr_{+}r_{-}a_{+}a_{-}s},\\
|\textbf{a}_-\rangle &=|N-1,0,0,0,1,0\rangle_{gr_{+}r_{-}a_{+}a_{-}s}.\\
\end{align*}
The Hamiltonian is closed in a subspace of Hilbert space that can be divided into the direct sum of two closed subspaces:
\begin{equation}
\mathcal {H}=(\mathcal {H}^1\otimes\mathcal {H}^2_{+})\oplus(\mathcal {H}^1\otimes\mathcal {H}^2_-).
\end{equation}
Here, $\mathcal {H}^1\otimes\mathcal{H}^2_{+}$ is spanned by basis vectors $\{|100\rangle$, $|\textbf{r}_+00\rangle$, $|001\rangle$, $|0\textbf{r}_+0\rangle$, $|0\textbf{a}_+0\rangle\}$, and
$\mathcal {H}^1\otimes\mathcal{H}^2_{-}$ is spanned by basis vectors $\{|110\rangle$, $|\textbf{r}_+10\rangle$, $|011\rangle$, $|0\textbf{r}_-0\rangle$, $|0\textbf{a}_-0\rangle\}$. The first two degrees of freedom represent cloud 1 and cloud 2, and the last degree of freedom of the state is the Fock state of the cavity photons. Since the Hamiltonian doesn't couple these two subspaces, and has the same form in each subspace, we can just consider a single subspace, e.g., $\mathcal {H}^1\otimes\mathcal {H}^2_+$. The Hamiltonian can be represented in this subspace as
\begin{equation}
\begin{split}
&H_3^{\textrm{eff}}(t)=\\
&\left[
\begin{array}{ccccc}
 0 & 0 & g_1 & 0 & \Omega_1\sin\theta \\
 0 & 0 & g_1 & \Omega_2\cos\theta e^{i\varphi_2} & 0 \\
 g_1 & g_1 & 0 & 0 & 0 \\
 0 & \Omega_2\cos\theta e^{-i\varphi_2} & 0 & 0 & 0 \\
 \Omega_1\sin\theta  & 0 & 0 & 0 & 0
\end{array}
\right],
\end{split}
\end{equation}
where we have set $\varphi_1=0$. A dark state exists for this system, since one eigenstate has eigenvalue 0:
\begin{equation}\label{darkstate}
\begin{split}
|D(t)\rangle &=\frac{\frac{g_1}{\Omega_1(t)}\cos\theta}{\sqrt{\frac{g_1^2}{\Omega_1^2(t)}\cos^2\theta+\frac{g_2^2(t)}{\Omega_2^2(t)} \sin^2\theta+ \cos^2\theta \sin^2\theta}}|100\rangle \\
&+\frac{e^{-i \varphi_2}\frac{g_2}{\Omega_2(t)}\sin\theta}{\sqrt{\frac{g_1^2}{\Omega_1^2(t)}\cos^2\theta+\frac{g_2^2(t)}{\Omega_2^2(t)} \sin^2\theta+ \cos^2\theta \sin^2\theta}}|0\textbf{a}_+0\rangle\\
&-\frac{\cos\theta \sin\theta}{\sqrt{\frac{g_1^2}{\Omega_1^2(t)}\cos^2\theta+\frac{g_2^2(t)}{\Omega_2^2(t)} \sin^2\theta+ \cos^2\theta \sin^2\theta}}|001\rangle.
\end{split}
\end{equation}
Just like the case of the phase gate, if the system is initially prepared in the dark state, after a cycle of adiabatic evolution a pure geometric phase can be obtained. In this case, if the initial
state is $|10\rangle_L$, we get a phase shift
\begin{equation}\label{phase_3}
\begin{split}
\phi_3&=i\oint d\textbf{R}\langle D(t)|\nabla_{\textbf{R}}|D(t)\rangle\\
        &=\oint d\varphi_2\frac{\frac{g_2^2}{\Omega_2^2(t)}\sin^2\theta}{\frac{g_2^2}{\Omega_2^2(t)}\sin^2\theta+\frac{g_1^2}{\Omega_1^2(t)}\cos^2\theta+\cos^2\theta\sin^2\theta}.
\end{split}
\end{equation}
Note that this phase is not affected by the spatially inhomogeneous distribution of the cavity mode.

We also get a geometric phase for the subspace $\mathcal {H}^1\otimes\mathcal {H}^2_-$.  However, we can set choose our path to set this phase to 0 independently of $\phi_3$. So, if we initially prepared the state $|10\rangle_L$ we get a geometric phase of $\phi_3$ after the gate manipulations, and if the initial state is $|11\rangle_L$ the geometric phase is 0. The states  $|00\rangle_L$ and $|01\rangle_L$ are not affected by this Hamiltonian, and hence also acquire no phase. The net effect is a two-qubit quantum gate,
\begin{align*}
&|00\rangle_L\rightarrow|00\rangle_L,\\
&|01\rangle_L\rightarrow|01\rangle_L,\\
&|10\rangle_L\rightarrow e^{i\phi_3}|10\rangle_L,\\
&|11\rangle_L\rightarrow|11\rangle_L,
\end{align*}
which is exactly the controlled phase gate we would like to implement.

\begin{figure}
\centering\includegraphics[width=80mm,height=60mm]{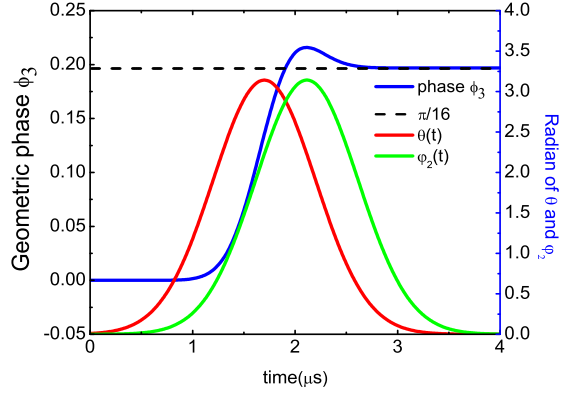}
\caption{\label{refer} The evolution of the phase $\phi_3$ of $|10\rangle_L$. The pulse shapes of $\theta(t)$ and $\varphi(t)$ are also given.}
\end{figure}

As we can see in Eq.~(\ref{darkstate}), the two clouds of atoms interact with each other by virtually absorbing and emitting a photon in cavity mode $a$, so leakage of cavity photons is the most important source of decoherence in our scheme. Analytically, the average photon number in the cavity during the process is
\begin{equation}
n_{\textrm{ph}+}= \frac{\cos^2\theta\sin^2\theta}{\frac{g_1^2}{\Omega_1^2(t)}\cos^2\theta+\frac{g_2^2}{\Omega_2^2(t)}\sin^2\theta+\cos^2\theta\sin^2\theta},
\end{equation}
when the system is initially prepared in $|10\rangle_L$, and
\begin{equation}
n_{\textrm{ph}-}= \frac{\cos^2\theta\sin^2\theta}{\frac{g_1^2}{\Omega_1^2(t)}\cos^2\theta+\frac{g_3^2}{\Omega_3^2(t)}\sin^2\theta+\cos^2\theta\sin^2\theta},
\end{equation}
when the system is initially prepared in $|11\rangle_L$, respectively.  If the initial state is $|00\rangle_L$ or $|01\rangle_L$ there are no photons in the cavity.

\begin{figure*}[!htp1]
\centering\includegraphics[width=120mm,height=55mm]{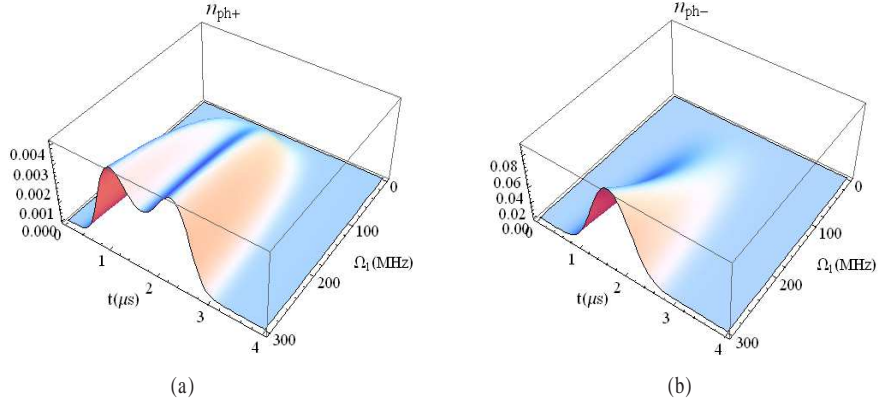}
\caption{\label{Fig:cavityphoton} The average number of photons in the cavity during gate operation as a function of time and $\Omega_1$ for $g_1=660$MHz. (a) system initially prepared in $|10\rangle_L$ and (b) system initially prepared in $|11\rangle_L$.}
\end{figure*}

Let's consider the concrete example of a conditional phase gate with $\phi_3=\pi/16$. We set the parameters to be $g_+ = 20\textrm{MHz}$, $g_3= g_-=10\textrm{MHz}$, and the number of atoms in each cloud to be $N=10^3$. Then we have $g_1=660\textrm{MHz}$, $\Omega_1= 40\textrm{MHz}$, $\Omega_2=50\textrm{MHz}$ and $\Omega_3=300\textrm{MHz}$. We choose
\[
\theta(t) = \frac{\pi}{4}\exp[-(t-1.7)^2/0.5] ,
\]
\[
\varphi_2(t) = \frac{\pi}{4}\exp[-(t-2.115)^2/0.5] ,
\]
(where time is expressed in $\mu$s). A numerical simulation (not including cavity loss) is shown in Fig.~\ref{refer} where the initial state is $\frac{1}{4}|00\rangle_L+\frac{1}{4}|01\rangle_L+\frac{1}{4}|10\rangle_L+\frac{1}{4}|11\rangle_L$. The whole process is very fast and can be finished in 4$\mu$s, with extremely high accuracy approaching the theoretical value calculated in Eq.~(\ref{phase_3}).

Fig.~\ref{Fig:cavityphoton} shows the average number of cavity photons $n_{\textrm{ph}+}$ and $n_{\textrm{ph}-}$ during the process as a function of time and $\Omega_1$ (not including cavity loss). The photon number is less than 0.001 in general which means that, the probability of a photon loss in the cavity is bounded above by 0.001 for the parameters given previously. We see that when the state of system is initially in $|11\rangle_L$, the average number of photons in the cavity is larger than in $|10\rangle_L$, $|00\rangle_L$ and $|01\rangle_L$. (Indeed, for the last two states the photon number is strictly zero.)

\begin{figure}[!hp]
\centering\includegraphics[width=70mm,height=60mm]{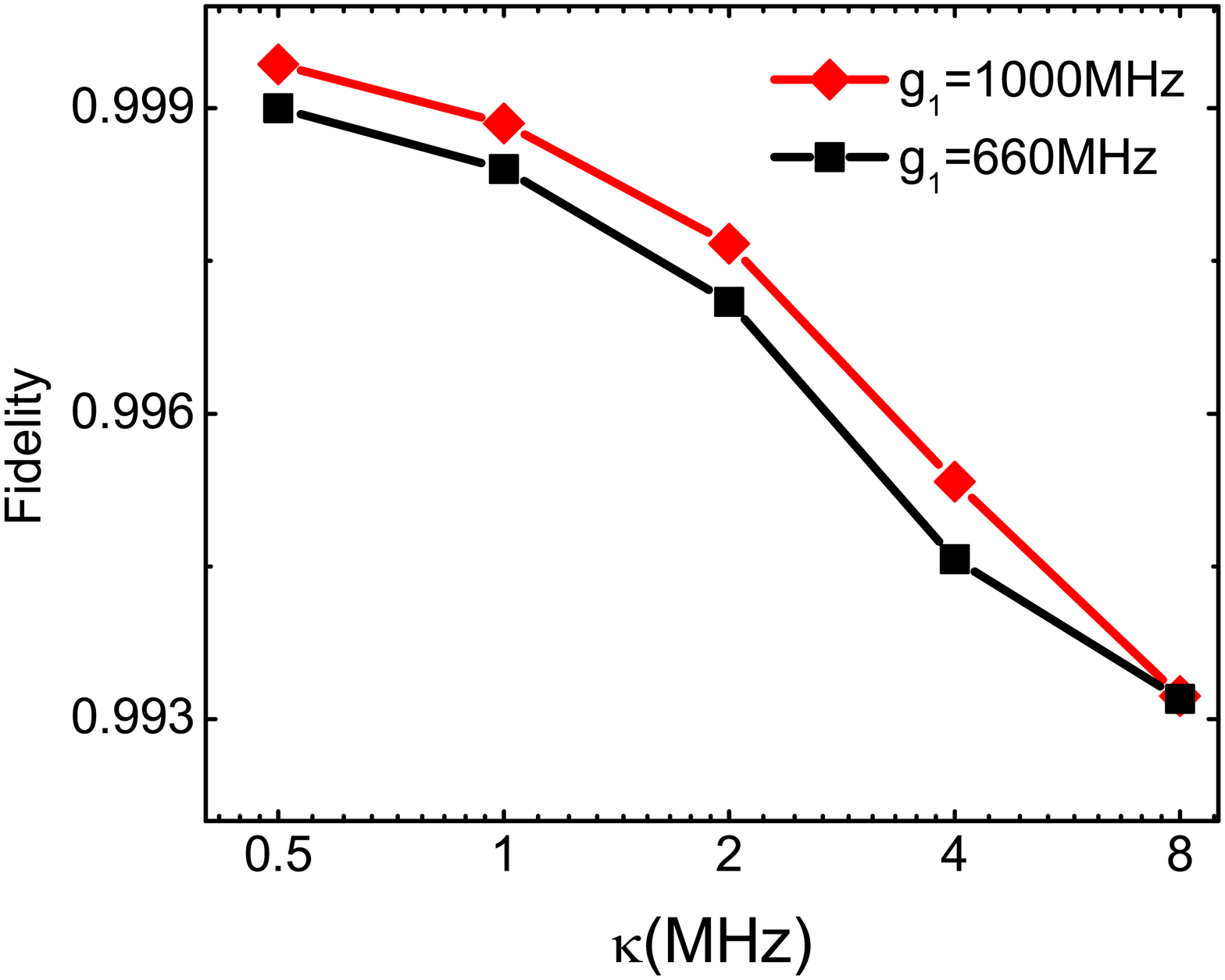}
\caption{\label{Fig:fidelity} Fidelity of the controlled phase gate with $\phi_3=\pi/16$ for different values of $\kappa$ and $g_1$.}
\end{figure}

Suppose the cavity loss rate is $\kappa$.  If that is the dominant decoherence channel, the fidelity of our conditional gate can be given by \cite{Nielson&Chuang}:
\begin{equation}\label{fidelity}
F=\min_{|\psi\rangle}\sqrt{\langle \psi|U^{\dagger}(\pi/16)\hat{\hat{\mathcal{L}}}(|\psi\rangle\langle \psi|)U(\pi/16)|\psi\rangle},
\end{equation}
where the superoperator $\hat{\hat{\mathcal{L}}}$ is determined by the master equation:
\begin{equation}
\frac{d\rho}{dt}=-i[H^{\textrm{eff}}_3,\rho]+2\kappa[a\rho a^\dagger-\frac{1}{2}\{a^\dagger a,\rho\}]
\end{equation}
in the Markov approximation. As stated previously, the initial state $|11\rangle_L$ emits more photons than the other three basis states during the gate operation. Hence, we can take $|\psi\rangle=|11\rangle_L$ as the worst case.

From Eq.~(\ref{phase_3}), we can see that  $\phi_3$ is approximately independent of $g_1$ and $g_2$ when they are large enough, so we can compare the fidelity for different value of $g_1$ for the same operation. Fig.~\ref{Fig:fidelity} shows the gate fidelity for different values of $\kappa$ and $g_1$ in our gate operation. The simulation is realized with C++ library ``Quantum State Diffusion" \cite{C++ Lib} using a fourth-order Runge-Kutta integrator \cite{Numerical Recipe} and a pseudo-random number generator to solve the master equation using quantum trajectories. (For each $\kappa$, we computed only 15 trajectories, so this is not a highly precise calculation.)  It makes sense that $g_1=1000\textrm{MHz}$ has better performance than that of $g_1=660\textrm{MHz}$, since the large value of $g_1$ decreases the number of cavity photons and hence gives a higher fidelity. The fidelity can exceed 0.999 when $\kappa=500\textrm{kHz}$ for $g_1=660\textrm{MHz}$, and when $\kappa=1\textrm{MHz}$ for $g_1=1000\textrm{MHz}$. For a FP cavity larger than $100\mu$m, values of $\kappa$ under $1\textrm{MHz}$ are quite achievable.

\section{Measurement}
 In fault-tolerant quantum computation, one must do syndrome measurement periodically \cite{Nielson&Chuang,Got97,Got98}, and the measurement results should be correct with high probability. It is also necessary to read out results at the end of a computation. Accurate measurement is therefore necessary. In this section, we propose a protocol to measure an atomic ensemble qubit in the computational basis.

Collecting the fluorescence from an atom is the most natural way to realize a measurement.  For a single atom (or ion), many fluorescence photons are needed to make the measurement reliable, so a ``cycling" transition from the computational state ($|0\rangle_L$ or $|1\rangle_L$) to an unstable excited state is generally used. In our case, however, there is an added complication.  The computational states include superpositions of a single excitation over all the atoms of the cloud.  The distance between atoms in the cloud can be several times larger than the wavelength of a photon emitted from the cloud, and this might make it possible (in principle) to distinguish which atom emitted the photon. This in turn could cause decoherence by collapsing the symmetric superposition, and take the system state outside the computational space. So we need a method to perform a reliable measurement while avoiding this problem.

\begin{figure}[!hp]
\vfill
\centering\includegraphics[width=80mm,height=50mm]{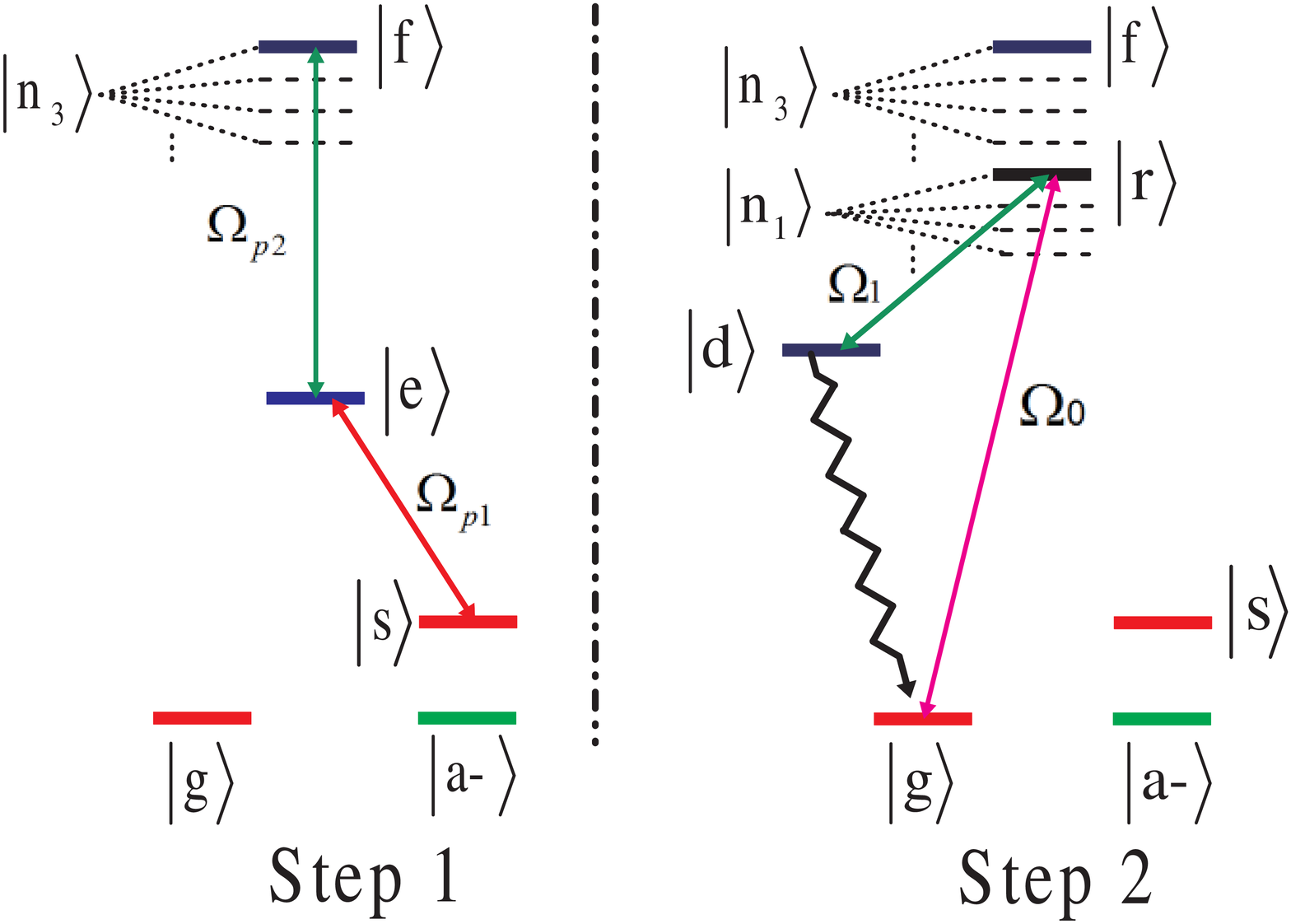}
\caption{\label{Fig:algorithm} The schematic setup to realize a projective measurement on a single qubit. Here, state $|d\rangle$ is a temporary state with a short lifetime. The transition between $|d\rangle$ and $|s\rangle$ is not allowed. The state $|d\rangle$ can be chosen for example, to be $|F=2,m_F=-2\rangle$ of the manifold $5{\rm P}_{3/2}$.}
\end{figure}

As shown in Fig. 13, the procedure can be realized in three steps:
\begin{enumerate}
  \item Adiabatically transport the population of $|1\rangle_L$ to $|{\bf f}\rangle$. (The procedure is similar to that used in realizing the gate $R_y(\phi)$.)
  \item Dynamically pump the atom cloud from the ground state to state $|{\bf r}\rangle$ with a $\pi$ pulse $\Omega_0$, and subsequently apply another $\pi$ pulse $\Omega_1$ between the state $|r\rangle$ and $|d\rangle$. Cycle this transition hundreds of times and collect the spontaneously emitted photons from the atom cloud.
  \item Reverse step 1 to transfer the population of $|{\bf f}\rangle$ back to $|1\rangle_L$ to prepare for next operation.
\end{enumerate}

If the qubit is initially in $|0\rangle_L$, step 1 has no effect.  The state of the cloud will transfer to $|{\bf r}\rangle$ and then to $|{\bf d}\rangle$ after two $\pi$ pulses; the atom cloud will quickly decay to the ground state and emit a photon.  By repeating this cycle many times, the probability of detecting the photons can be made quite high.

If the qubit is initially in $|1\rangle_L$, step 1 will produce a single excitation in $|f\rangle$, and by the dipole-dipole blockade effect the transition from the ground state to $|r\rangle$ will be blocked. Therefore we will not observe any emitted photons. Thus we observe fluorescence only if the qubit is initially in the state $|0\rangle_L$, and this procedure is a projective measurement. Note that non-uniform coupling is not a problem here since at the end of the procedure the state will be either $|0\rangle_L$ or $|1\rangle_L$.

The time for a single $\pi$ laser pulse (and the accompanying state transition) should be roughly 10-20ns when Rabi frequency of the pumping laser is hundreds of MHz. Assuming a life time of about 10ns for $|d\rangle$, each cycle of measurement can be finished in 20-30ns. After the cycling transitions, we must transfer the population of $|{\bf f}\rangle$ back to $|1\rangle_L$, so we need the total measurement time to be less than the lifetime of $|f\rangle$ (estimated to be about 400$\mu$s).  This gives an upper limit on the total number of cycles we can make during the process. If we want to limit the error to a reasonable level, we can perform at most a few hundred transition cycles. This should be adequate if the photodetector has sufficiently high efficiency and large enough solid angle of detection.

\section{Conclusions}

In this paper we have proposed a scheme to achieve quantum computation using geometric manipulation of ensembles of atoms in a cavity QED system. Adiabatic optical control can be used to obtain a geometric phase gate and controlled phase gate. Combining optical excitation with a dipole-dipole blockade between Rydberg states allows us to geometrically realize the $R_y(\phi_2)$ gate. Thus a universal set of quantum gates can  be realized geometrically.

We analyzed this scheme for ensembles of neutral rubidium atoms, magnetically trapped in planoconcave microcavities on an atom chip.  Numerical simulations show that a single qubit gate can be performed in several microsecond with very low probability of gate error.  For the controlled phase gate, the operation is done by virtually emitting and absorbing a photon from the cavity mode, and can be completed in 4$\mu$s for a controlled $\pi/32$ gate. An advantage of geometric manipulation is that by adiabatic parameter control we can avoid certain kinds of errors, especially those caused by inhomogeneous distribution of the laser beam and cavity modes. The values of the fields can depend on atom position, but their ratio can be fixed and controllable. The basic idea is to find an appropriate adiabatic process, so that the relevant dynamics are either independent of, or depend only on the ratio of, the two coupling rates.

The ensemble of atoms effectively enlarges the coupling rate $g$ by $\sqrt{N}$, which greatly suppresses the likelihood of cavity photons and increases the fidelity of the operation.  We analyzed the scheme for $N=10^3$ atoms in each cloud.  The parameters that we have assumed in our numerical simulations are all achievable by current experiments.  This, together with the possibility of coupling stationary qubits (for computation) with flying qubits (for communication) makes this scheme look particularly promising for near-term quantum protocols.

\section*{ACKNOWLEDGMENTS}

TAB and YZ acknowledge support from NSF Grants EMT-0829870 and CCF-0448658.

\section*{APPENDIX}

In this appendix, we discuss how to choose the appropriate Stark eigenstates of alkali atoms for our purpose. Since alkali atoms have spectra similar to hydrogen, we first analyze the effect of the dipole-dipole interaction of Rydberg stark eigenstates for hydrogen. After that, we allow for the difference between alkali atoms and hydrogen, and discuss a method of calculating their energy structure. Using this method, we will choose appropriate Rydberg Stark eigenstates of rubidium as an example.  (Please note that the fine structure of the p and d states of rubidium have observable effects on the spectrum. However, for simplicity, we do not take this into account.) The method of calculation is described in detail in Ref. \cite{Rydberg atoms}. Note that all quantities below are in atomic units for simplicity.

We first consider Stark states of a single hydrogen atom. The magnetic quantum number $m$ is a good quantum number. From perturbation theory, the first order approximation of Stark eignstates are parabolic states $|n,n_1,n_2,m\rangle$ \cite{Rydberg atoms} with energies
\[
E_{n,n_1,n_2,m} = - \frac{1}{2n^2} + \frac{3}{2}E (n_1 - n_2)n .
\]
Here, $n$ is the principal quantum number, $E$ is the electric field in the $z$ direction, and $n_1$, $n_2$ are non-negative integers satisfying the equality $n=n_1 + n_2 +|m|+1$.  For $m=0$, allowed values of $n_1 - n_2$ are $n-1,n-3,...,-n+1$ and for $m = 1$, they are $n-2,n-4,...,-n+2$. In the following discussion, we use the quantum number $q=n_1 - n_2$  instead of $n_1$ and $n_2$ for simplicity.

We expand the Stark parabolic states $|n,m,q\rangle$ in the spherical basis $|n,l,m\rangle$ as \cite{Rydberg atoms}:
\begin{equation}
|n,m,q\rangle = \sum_{l=0}^{n-1}|n,l,m\rangle\langle n,l,m|n,m,q\rangle.
\end{equation}
The coefficients can be written in terms of Wigner 3J symbols \cite{Rydberg atoms}:
\begin{equation}
\begin{split}
c(n,l,m,q)\equiv&\langle n,l,m|n,m,q\rangle = (-1)^{(1-n+m+q)/2+l} \\
&\times\sqrt{2l+1}\left(
                    \begin{array}{ccc}
                      \frac{n-1}{2} & \frac{n-1}{2} & l \\
                      \frac{m+q}{2} & \frac{m-q}{2} & -m \\
                    \end{array}
                  \right).
\end{split}
\end{equation}
For example, if $n=2$, $m=0$, $q=1$, we have $c(2,0,0,1)=\sqrt{2}/2$ and $c(2,1,0,1)=-\sqrt{2}/2$. So, $|2,0,1\rangle=(\sqrt{2}/2)|2,0,0\rangle-(\sqrt{2}/2)|2,1,0\rangle$. We will frequently use these parabolic states instead of spherical states in the following analysis.

The dipole-dipole interaction $V_{\textrm{dip}}$ is proportional to  $\widehat{\textbf{r}}_1\cdot\widehat{\textbf{r}}_2$, where
\[
\widehat{\textbf{r}}_1=r_1(\frac{x_1}{r_1}\textbf{e}_x+\frac{y_1}{r_1}\textbf{e}_y+\frac{y_1}{r_1}\textbf{e}_z)\otimes I ,
\]
\[
\widehat{\textbf{r}}_2=I\otimes r_2(\frac{x_2}{r_2}\textbf{e}_x+\frac{y_2}{r_2}\textbf{e}_y+\frac{y_2}{r_2}\textbf{e}_z) ,
\]
are coordinate operators for atoms 1 and atom 2. Replacing $x$, $y$, $z$ with spherical coordinates $r$, $\theta$, $\phi$,
\begin{equation}
\begin{split}
x&=r\sin\theta\cos\phi,\\
y&=r\sin\theta\sin\phi,\\
z&=r\cos\theta.
\end{split}
\end{equation}
gives us
\begin{equation}
\begin{split}
&V_{\textrm{dip}}\propto\widehat{\textbf{r}}_1\cdot\widehat{\textbf{r}}_2\\ =
 &r_1(\frac{x_1}{r_1}\textbf{e}_x+\frac{y_1}{r_1}\textbf{e}_y+\frac{y_1}{r_1}\textbf{e}_z)\cdot r_2(\frac{x_2}{r_2}\textbf{e}_x+\frac{y_2}{r_2}\textbf{e}_y+\frac{y_2}{r_2}\textbf{e}_z)\\
=&r_1\sqrt{\frac{4\pi}{3}}( \frac{\textbf{e}_x+i\textbf{e}_y}{\sqrt{2}}Y_{1,-1}^{1}+ \frac{i\textbf{e}_y-\textbf{e}_x}{\sqrt{2}}Y_{1,1}^{1}+Y_{1,0}^1\textbf{e}_z)\cdot\\
&r_2\sqrt{\frac{4\pi}{3}}( \frac{\textbf{e}_x+i\textbf{e}_y}{\sqrt{2}}Y_{1,-1}^{2}+ \frac{i\textbf{e}_y-\textbf{e}_x}{\sqrt{2}}Y_{1,1}^{2}+Y_{1,0}^2\textbf{e}_z)\\
=&\frac{4\pi}{3}r_1 r_2 (- Y_{1,-1}^1Y_{1,1}^2 - Y_{1,1}^1 Y_{1,-1}^2 + Y_{1,0}^1Y_{1,0}^2),
\end{split}
\end{equation}
where $Y_{l,m}^{i}$ is a spherical harmonic function for atom $i$. Generally speaking, for two atoms in the given initial Stark eigenstate,  the diagonal terms of the dipole-dipole interaction give an energy shift, while its non-diagonal terms couple adjacent $m$ manifolds with each other: $(m, m)$ to $(m\pm1,m\mp1)$. The Stark states that are most useful for our scheme are those that maximize the energy shift while suppressing the transition between different $m$ manifolds (which might introduce decoherence channels). So, it is sufficient for us to know the value of the matrix elements
\begin{equation}
\begin{split}
\langle n,m|\langle n,m,q |V_{\textrm{dip}}|n^\prime,m+1\rangle|n^\prime, m-1\rangle,\\
\langle n,m|\langle n,m,q |V_{\textrm{dip}}|n^\prime,m-1\rangle|n^\prime, m+1\rangle,
\end{split}
\end{equation}
and
\begin{equation}\label{eq:stark_eigen}
\langle n,m,q|\langle n,m,q |V_{\textrm{dip}}|n^\prime,m,q\rangle|n^\prime, m,q\rangle.
\end{equation}
For hydrogen, we will only consider matrix elements with the same $n$ for simplicity. Even so, it is too complicated (and unnecessary) to find the general analytical form of these elements. Instead, we numerically do the integration to determine which states fulfill our requirements. By symmetry, the transition strengths of $(m,m)\rightarrow(m+1,m-1)$ and $(m,m)\rightarrow(m-1,m+1)$ are the same, so we just give the first value. The calculation results of some matrix elements are shown in Table.~\ref{table1} as an example.

\begin{table*}
\begin{center}
\begin{tabular}{lccccccc}
\hline
Matrix element:  $|m, q\rangle\rightarrow|m^\prime, q^\prime\rangle$& n  =5 & n = 10 &n = 15 &n = 20 &n = 25\\
\hline
$|0,n-1\rangle|0,n-1\rangle\rightarrow |1,n-2\rangle|-1,n-2\rangle$& 112.5 & 1012.5 & 3543.75&8550&16875\\
$|0,n-1\rangle|0,n-1\rangle\rightarrow |0,n-1\rangle|0,n-1\rangle$&900 & 18225 & 99225&324900&810000\\
$|0,n-1\rangle|0,n-1\rangle\rightarrow |1,n-4\rangle|-1,n-4\rangle$& 0 & 0 & 0&0&0\\
\hline
$|0,n-3\rangle|0,n-3\rangle\rightarrow |1,n-2\rangle|-1,n-4\rangle$ & 137.78 & 1350 &4829.32&11769&23362.4\\
$|0,n-3\rangle|0,n-3\rangle\rightarrow |1,n-4\rangle|-1,n-2\rangle$ & 137.78 & 1350 &4829.32&11769&23362.4\\
$|0,n-3\rangle|0,n-3\rangle\rightarrow |1,n-4\rangle|-1,n-4\rangle$ & 168.75 & 1800 &6581.25&16200&32343.7\\
$|0,n-3\rangle|0,n-3\rangle\rightarrow |1,n-2\rangle|-1,n-2\rangle$ &112.5 & 1012.5 & 3543.75&8550&16875\\
$|0,n-3\rangle|0,n-3\rangle\rightarrow |0,n-3\rangle|0,n-3\rangle$ &225.01 & 11025 & 72900 &260100&680625\\
\hline
$|1,n-2\rangle|1,n-2\rangle\rightarrow |2,n-3\rangle|0,n-3\rangle$ & 137.78 & 1350 & 4829.32 &11769 &23362.4\\
$|1,n-2\rangle|1,n-2\rangle\rightarrow |1,n-2\rangle|1,n-2\rangle$ & 506.25  &14400 & 85556&291600&743906\\
\hline
$|1,n-4\rangle|1,n-4\rangle\rightarrow |2,n-3\rangle|0,n-5\rangle$ & 137.78 & 1350 & 4829.32&11769&23362.4\\
$|1,n-4\rangle|1,n-4\rangle\rightarrow |2,n-5\rangle|0,n-3\rangle$ & 168.75 & 2062.16& 7744.14&19281.9 &38742.1\\
$|1,n-4\rangle|1,n-4\rangle\rightarrow |2,n-5\rangle|0,n-5\rangle$ & 168.75 & 2062.16 & 7744.14 &19281.9 &38742.1\\
$|1,n-4\rangle|1,n-4\rangle\rightarrow |2,n-3\rangle|0,n-3\rangle$ & 137.78 & 1350 & 4829.32&11769&23362.4\\
$|1,n-4\rangle|1,n-4\rangle\rightarrow |1,n-4\rangle|1,n-4\rangle$ & 56.25  & 8100 & 61256.3&230400&620156\\
\hline
\end{tabular}
\caption{The matrix element of operator $V_{\rm{dip}}$.}\label{table1}
\end{center}
\end{table*}

We have made several simplifications in this table. First, we didn't calculate the transition strengths between different $n$ manifolds, because they are several times smaller than their counterparts in the same $n$ manifold. Second, we didn't calculate the case when two atoms are initially prepared in different Stark eigenstates, especially in two different manifolds, which actually we {\it have} proposed to realize both the gate $e^{\phi_2\sigma_y}$ and qubit measurement. Nevertheless, this table gives enough information about the characteristics of the state we are looking for. First, the transition strength must be much smaller than the energy shift inside the same $n$ manifold. Second, those transitions between initial and final states with a difference in parabolic number $q$ larger than one must be greatly suppressed. If we prepare an atom in the outermost state $|n,m=0,q=n-1\rangle$, we obtain the largest energy shift with the smallest transition strength to the $(m+1,m-1)$ state compared with other Stark eigenstates in the same manifold.

In our proposed gate, since the Rydberg states of two atoms may not be in the same manifold, a natural solution to fulfill our requirements is to choose state $|\Psi\rangle=|n,m=0,q=n-1\rangle|n^{\prime},m=0, q= n^{\prime}-1\rangle$ for manifolds $n$ and $n^{\prime}$. Note that the energy shift term in the Hamiltonian is a direct product of operators on two different atoms, so the analysis of two atoms in same manifold can be directly applied to show that the maximum energy shift is obtained if the two atoms are prepared in the state $|\Psi\rangle$.  This gives an energy shift of
\[
\langle \Psi|V_{\textrm{dip}}|\Psi\rangle \propto n(n-1)n^\prime(n^\prime-1) .
\]

Next, we consider the case of non-hydrogen alkali atoms. Physically, the main difference between alkali atoms and hydrogen atoms is that the former have a finite sized ionic core that results in avoided crossings in the Stark spectrum. So, the Hamiltonian can be written as
\begin{equation}
H=-\frac{\nabla^2}{2}-\frac{1}{r}+V_d(r)+Ez.
\end{equation}
Here, $V_d(r)$ is the difference of the potential function between a hydrogen and an alkali atom, due to the finite-sized ionic core. We treat $V_d(r)$ as spherically symmetric and only nonzero near the nucleus. In the case where quantum defects are relatively small (e.g., $n$ is large), we can use the hydrogenic parabolic states as our working basis. For diagonal terms of the Hamiltonian, we have:
\begin{equation}
\begin{split}
&\langle n,m,q|H|n,m,q\rangle = -\frac{1}{2n^2}+\frac{3}{2}qnE\\
 & +\langle n,m,q|V_d(r)|n,m,q\rangle+O(E^2),
\end{split}
\end{equation}
and we represent the non-diagonal terms using hydrogen spherical states $|n,l,m\rangle$:
\begin{equation}
\begin{split}
&\langle n,m,q|H|n^{\prime},m,q^{\prime}\rangle= \sum_{l,l^{\prime}}^{n-1}\langle n,m,q|n,l,m\rangle \\
&\times\langle n,l,m|V_d(r)|n^{\prime},l^{\prime},m\rangle \langle n^{\prime},l^{\prime},m|n^{\prime},m,q^{\prime}\rangle.
\end{split}
\end{equation}

\begin{figure*}[!ht]
\centering\includegraphics[width=130mm,height=100mm]{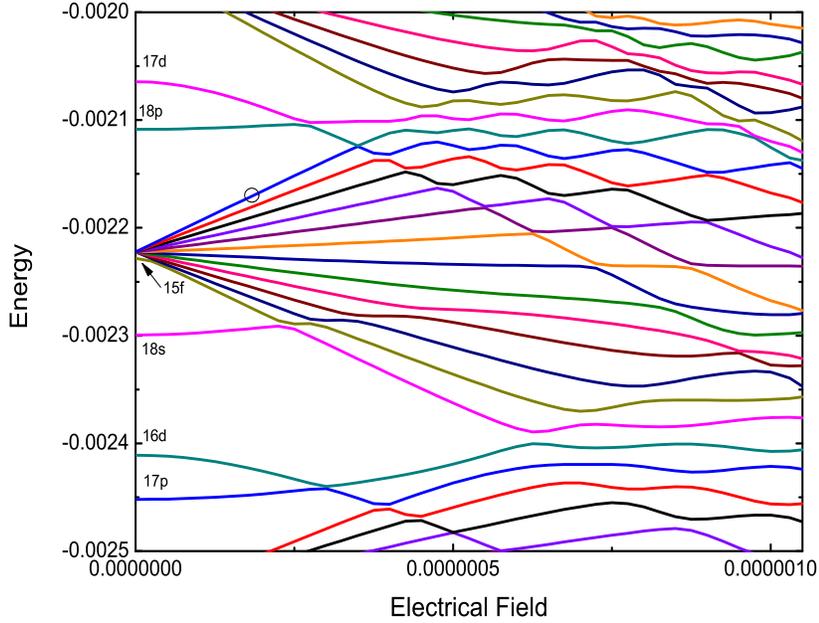}
\caption{\label{Fig:stark}The energy level of rubidium around manifold $n=15$, $m=0$. The circled states is our candidate states which have large energy shift.}
\end{figure*}

Denote the spherically symmetric eigenstates of the alkali atom as $|n,l,m\rangle_{al}$. For large $n$, we have
\begin{equation}
\begin{split}
-\frac{1}{2(n-\delta_l)^2}=&_{al}\langle n,l,m|H|n,l,m\rangle_{al} \\
\approx & \langle n,l,m|\left(-\frac{\nabla^2}{2}-\frac{1}{r}+V_d(r)\right)|n,l,m\rangle\\
=&-\frac{1}{2n^2}+\langle n,l,m|V_d(r)|n,l,m\rangle.
\end{split}
\end{equation}
where $\delta_l$ is the quantum defect for angular momentum $l$ of the alkali atom \cite{Atomic Physics}, which is different for each element. For rubidium, if we neglect neglect the fine structure effect, $\delta_0\approx 3.1$, $\delta_1\approx 2.6$, $\delta_2\approx 1.3$, and $\delta_3\approx 0.02$. For $l>3$, $\delta_l \approx 0$. On the other hand,
\[
-\frac{1}{2(n-\delta_l)^2}\approx -\frac{1}{2n^2} - \frac{\delta_l}{n^3} .
\]
So we have:
\begin{equation}
\langle n,l,m|V_d(r)|n,l,m\rangle\approx -\frac{\delta_l}{n^3}.
\end{equation}
Per Ref. \cite{Rydberg atoms}, this expression may be generalized to
\begin{equation}
\langle n,l,m|V_d(r)|n^{\prime},l,m\rangle = -\frac{\delta_l}{\sqrt{n^{3}n^{\prime 3}}}.
\end{equation}
Here, we take advantage of the spherical symmetry of $V_d(r)$, so the matrix element vanishes when the $l$ value on the two sides of the above equation are not the same.

Observe that Eq.~(\ref{eq:stark_eigen}) can be represented as matrix multiplication. We define matrices
\begin{equation}
\begin{split}
[\mathcal{C}^{nm}]_{ql} =& \langle n,m,q|n,l,m\rangle=c(n,l,m,q) ,\\
[\mathcal{V}^{nm}]_{lq} =& \langle n,l,m|n,m,q\rangle=c(n,l,m,q) ,\\
[\mathcal{D}^{nn^{\prime}}]_{ll^{\prime}} =& \langle n,l,m|V_d(r)|n^{\prime},l^{\prime},m\rangle , \\
[\mathcal{S}^{nn^{\prime}m}]_{qq^{\prime}} =& \langle n,m,q|H|n^{\prime},m,q^{\prime}\rangle .
\end{split}
\end{equation}
Then, we have
\begin{equation}
\mathcal{S}^{nn^{\prime}m} = \mathcal{C}^{nm} \mathcal{D}^{nn^{\prime}} \mathcal{V}^{n^{\prime}m}.
\end{equation}
For the purpose of illustration, we show a simple example. We write the matrices
\begin{equation*}
\mathcal{C}^{20}=\left[
\begin{array}{cc}
 \frac{1}{\sqrt{2}} & \frac{1}{\sqrt{2}} \\
 \frac{1}{\sqrt{2}} & -\frac{1}{\sqrt{2}}
\end{array}
\right],
\end{equation*}
\begin{equation*}
\mathcal{V}^{30}=\left[
\begin{array}{ccc}
 \frac{1}{\sqrt{3}} & \frac{1}{\sqrt{3}} & \frac{1}{\sqrt{3}} \\
 \frac{1}{\sqrt{2}} & 0 & -\frac{1}{\sqrt{2}} \\
 \frac{1}{\sqrt{6}} & -\sqrt{\frac{2}{3}} & \frac{1}{\sqrt{6}}
\end{array}
\right],
\end{equation*}
\begin{equation*}
\mathcal{D}^{23}=\left[
\begin{array}{ccc}
 -\frac{\delta _0}{6 \sqrt{6}} & 0 & 0 \\
 0 & -\frac{\delta _1}{6 \sqrt{6}} & 0
\end{array}
\right],
\end{equation*}
and multiply them to get
\begin{equation*}
\mathcal{S}^{230}=\left[
\begin{array}{ccc}
 -\frac{\delta _0}{36}-\frac{\delta _1}{12 \sqrt{6}} & -\frac{\delta _0}{36} & -\frac{\delta _0}{36}+\frac{\delta _1}{12 \sqrt{6}} \\
 -\frac{\delta _0}{36}+\frac{\delta _1}{12 \sqrt{6}} & -\frac{\delta _0}{36} & -\frac{\delta _0}{36}-\frac{\delta _1}{12 \sqrt{6}}
\end{array}
\right].
\end{equation*}
Matrix $\mathcal{S}^{nn^{\prime}m}$ can be treated as a submatrix (or block) of the Hamiltonian matrix. Since the Hilbert space is infinite-dimensional, we need to truncate the Hamiltonian matrix.
For example, if we need to know eigenvalues and eigenstates of manifold $n$ of rubidium, we need to consider only those manifolds that couple to it.

We now calculate the energy structure of the $n=15$ manifold for the purpose of illustration. We also need to consider the adjacent manifolds such as $n=14$, $n=16$, $n=17$ and $n=18$ since states like  16d, 17p, 17d, 18s and 18p are coupled to the $n=15$ manifold. Define the submatrix of the Hamiltonian $\mathcal{H}^{nn^{\prime}m}$ as
\[
[\mathcal{H}^{nn^{\prime}m}]_{qq^{\prime}}=[\mathcal{S}^{nn^{\prime}m}]_{qq^{\prime}}-\delta_{nn^\prime}\delta_{qq^\prime}(\frac{1}{2n^2}-\frac{3}{2}qnE) ,
\]
where, $\delta_{nm}$ is  the Kronecker delta.  The Hamiltonian matrix is approximately represented as:
\begin{equation}
\left[
  \begin{array}{ccccc}
   \mathcal{H}^{14,14,m} & \mathcal{H}^{14,15,m} & \mathcal{H}^{14,16,m} & \mathcal{H}^{14,17,m} & \mathcal{H}^{14,18,m} \\
   \mathcal{H}^{15,14,m} & \mathcal{H}^{15,15,m} & \mathcal{H}^{15,16,m} & \mathcal{H}^{15,17,m} & \mathcal{H}^{15,18,m} \\
   \mathcal{H}^{16,14,m} & \mathcal{H}^{16,15,m} & \mathcal{H}^{16,16,m} & \mathcal{H}^{16,17,m} & \mathcal{H}^{16,18,m} \\
   \mathcal{H}^{17,14,m} & \mathcal{H}^{17,15,m} & \mathcal{H}^{17,16,m} & \mathcal{H}^{17,17,m} & \mathcal{H}^{17,18,m}\\
   \mathcal{H}^{18,14,m} & \mathcal{H}^{18,15,m} & \mathcal{H}^{18,16,m} & \mathcal{H}^{18,17,m} & \mathcal{H}^{18,18,m} \\
  \end{array}
\right],
\end{equation}
which is an $80\times80$ Hermitian matrix. Similar methods could be applied to much higher excitation states like $n=80$.

By diagonalizing the Hamiltonian matrix, we can obtain the spectrum and the eigenstates. As mentioned earlier, the state we use should have a large component of $|n,m=0,q=n-1\rangle$, which means we want the single-atom eigenstate
\[
|r\rangle_{n,m}=\sum_{q}C_q|n,m,q\rangle
\]
to have $m=0$ with a large coefficient $|C_q|$ for $q=n-1$.

Fig.~\ref{Fig:stark} shows the spectrum around the manifold $n=15$. The outermost Stark eigenstates $|r\rangle_{15,0}$, circled in the figure in the regime where the Stark eigenvalues are roughly linear in the electric field ($E\leq 2.5\times10^{-7}$), are good candidate states, since they should have behavior similar to the hydrogen state $|n,m=0,q=n-1\rangle$. Numerical calculation shows that the diagonal term of the dipole-dipole interaction is about $8\times10^{4}$ in $|r\rangle_{15,0}|r\rangle_{15,0}$ (compared to 99225 for hydrogen in the previous table), which is larger than for the other states in the manifold.

To put this in the context of our scheme for quantum computation, consider the case where there are $10^3$ atoms in a volume of $(6\mu$m$)^3$. The smallest energy shift of a pair of atoms inside the cell is for those that are most distant. Suppose the two most distant atoms are in the state $|r\rangle_{15,0}|r\rangle_{15,0}$. The energy shift between them should be around 1MHz, depending on the spatial distribution of atoms.  In our scheme, where two atoms may also be in different single-atom states (actually in different manifolds), we could naturally extend our original analysis, but we expect a similar result.

In practice, we would want to use higher energy Rydberg states; for instance an $n=60$ state $|m\rangle=|r\rangle_{60,0}$ for the intermediate state and an $n=70$ state $|r\rangle=|r\rangle_{70,0}$ for the register state, to get both longer lifetimes (more than 300-400$\mu$s for $n=70$ can be achieved in a cryogenic environment \cite{cryogenic}) and a stronger dipole-dipole interaction. Since higher energy Rydberg states should behave more like the Rydberg states of hydrogen, we can pick the outermost Stark eigenstates in the linear Stark area for both atoms, whether or not they have the same principal quantum number. If the initial state of one atom is in the outermost state of manifold $n=60$, and the other in either manifold $n=60$ or $n=70$, then by $V_{\textrm{dip}}\propto n(n-1)n^{\prime}(n^\prime-1)$ the energy shift should be roughly 200-300MHz or 300-400MHz, respectively. The distance of the most closely-spaced pair of atoms should be about $10^4a_0$.  This is larger than $2R$, where $R=4900a_0$ is the radius of atoms in manifold $n=70$. This spacing should maintain the validity of our dipole-dipole interaction model.


\begin{thebibliography}{}
\bibitem{Bennett}
D. P. Divincenzo, Science {\bf 270}, 255 (1995); C. H. Bennett,
Phys. Today {\bf 48}(10), 24 (1995); C. H. Bennett and D. P.
Divincenzo, Nature (London) {\bf 404}, 247 (2000).

\bibitem{Nielson&Chuang}
 M. A. Nielsen and
I. L. Chuang , {\it Quantum Computation and Quantum Information},
(Cambridge University Press, Cambridge 2000).

\bibitem{Quantum Computing Devices}
G. Chen {\it et al.}, {\it Quantum Computing Devices: Principles,
Devices, and Analysis},
 (Chapman \& Hall/CRC, 2006).


\bibitem{Ions trap_1}
J. I. Cirac and P. Zoller, Phys. Rev. Lett. {\bf 74}, 4091 (1995).

\bibitem{Ions_trap_2}
A. S{\o}rensen, K. M{\o}lmer, Phys. Rev. Lett. {\bf 82}, 1971
(1999).
\bibitem{Neutral_atom}
G. K. Brennen {\it et al.}, Phys. Rev. Lett. {\bf 82}, 1060 (1999).

\bibitem{Neutral_atom2}
D. Jaksch {\it et al.}, Phys. Rev. Lett. {\bf 85}, 2208 (2000).

\bibitem{Cavity_QED}
T. Pellizzari, S. A. Gardiner, J. I. Cirac and P. Zoller, Phys. Rev.
Lett. {\bf 75}, 3788 (1995).

\bibitem{Cavity_QED_Guo}
S.-B. Zheng, G.-C. Guo, Phys. Rev. Lett. {\bf 85}, 2392 (2000).

\bibitem{LOQC}
E. Knill, R. Laflamme, G. J. Milburn, Nature {\bf 409}, 46 (2001).

\bibitem{Lukin_EIT_Review}
M. D. Lukin, Rev. Mod. Phys. {\bf 75}, 457 (2003);

\bibitem{EIT_Review}
M. Fleischhauer, A. Imamo\v{g}lu, and J. P. Marangos, Rev. Mod.
Phys. {\bf 77}, 633 (2005).

\bibitem{EIT_Experiment_2009}
U. Schnorrberger {\it et al.}, Phys. Rev. Lett. {\bf 103}, 033003 (2009)

\bibitem{MDLukin2001}
M. D. Lukin {\it et al.}, Phys. Rev. Lett. {\bf 87}, 037901 (2001).

\bibitem{Photon_Photn_Interaction_2011}
A. V. Gorshkov, J. Otterbach, M. Fleischhauer, T. Pohl, M. D. Lukin, arXiv:1103.3700.

\bibitem{Muller2009}
M. M\"{u}ller, I. Lesanovsky, H. Weimer, H. P. B\"{u}chler, and P. Zoller,
Phys. Rev. Lett. {\bf 102}, 170502 (2009).

\bibitem{Weimer2010}
H. Weimer, M. M\"{u}ller, I. Lesanovsky, P. Zoller, H. P. B\"{u}chler,
Nature. Phys. {\bf 6}, 382 (2010).

\bibitem{Gaetan2009}
A. Ga\"{e}tan. {\it et al.}, Nature. Phys. {\bf 5}, 115 (2009).

\bibitem{Urban2009}
E. Urban {\it et al.}, Nature. Phys. {\bf 5}, 110 (2009).

\bibitem{Petresyan}
D. Petrosyan, M. Fleischhauer, Phys. Rev. Lett. {\bf 100}, 170501
(2008).

\bibitem{Folman2002}
R. Folman, P. Kr\"{u}ger, J. Schmiedmayer, J. Denschlag, and C.
Henkel, Adv. Atom. Mol. Phy. {\bf 48}, 263 (2002)

\bibitem{Review_of_Modern_Physics_Atomschip}
J. Fort\'{a}gh and C. Zimmermann, Rev. Mod. Phys. {\bf 79}, 235
(2007).

\bibitem{Trupke2007}
M. Trupke, J. Metz, A. Beige, E. A. Hinds, J. Mod. Opt. {\bf 54},
1639 (2007).

\bibitem{Trupke2005}
M. Trupke {\it et al.}, Appl. Phys. Lett. {\bf 87}, 211106 (2005).

\bibitem{Phase_gate_atom chips}
E. Charron {\it et al.}, Phys. Rev. A {\bf 74}, 012308 (2006).



\bibitem{Treutlein2004}
P. Treutlein {\it et al.}, Phys. Rev. Lett. {\bf 92} 203005(2004).

\bibitem{Trap_photon}
M. Fleishhauer, S. F. Yelin, M. D. Lukin, Opt. Comm. {\bf 179}, 395
(2000).

\bibitem{Trap_correlated_photon}
M. D. Lukin, S. F. Yelin, M. Fleischhauer, Phys. Rev. Lett. {\bf
84}, 4232 (2000).

\bibitem{Zanardi}
P. Zanardi, M. Rasseti, Phys. Lett. A {\bf 264}, 94 (1999).

\bibitem{Berry}
M. V. Berry, Proc. R. Soc. London A. {\bf 392}, 45 (1984).

\bibitem{Wilczek}
F. Wilczek, A. Zee, Phys. Rev. Lett. {\bf 52}, 2111 (1984).

\bibitem{Llyod}
S. Llyod, Phys. Rev. Lett. {\bf 75}, 346 (1995).

\bibitem{L-M-Duan_Science_2001}
L.-M. Duan, J. I. Cirac, P. Zoller, Science {\bf 292}, 1695 (2001).


\bibitem{cryogenic}
M. Saffman and T. G. Walker, Phys. Rev. A {\bf 72}, 022347(2005).

\bibitem{Swap}
J. Simon {\it et al.}, Nature.Phys. {\bf 3}, 765(2007)

\bibitem{FORT}
L.-M. Duan, A. Kuzimich, H. J. Kimble, Phys. Rev. A {\bf 67}, 032305(2003).

\bibitem{Quantum Optics}
D. F. Walls and G. J. Milburn, {\it Quantum Optics} (Springer-Verlag, 1994).


\bibitem{C++ Lib}
R. Shack and T. A. Brun, Comp. Phys. Comm. {\bf 102}, 210 (1997).

\bibitem{Numerical Recipe}
W. H. Press {\it et al.}, {\it Numerical Recipes in C} (Cambridge
University Press, 2002).

\bibitem{Got97}
D. Gottesman, Ph.D. thesis, Caltech, 1997.

\bibitem{Got98}
D. Gottesman, Phys. Rev. A {\bf 57}, 127 (1998).

\bibitem{Rydberg atoms}
T. F. Gallagher, {\it Rydberg Atoms}(Cambridge University Press, 1994).

\bibitem{Atomic Physics}
C. J. Foot, {\it Atomic Physics}(Oxford University Press, 2005).



\end{thebibliography}
\end{document}